\documentclass[trackchanges, twocolumn]{aastex701}

\usepackage{amsmath}

\shorttitle{TDEs on FIRE}
\shortauthors{Kar Chowdhury et al.}

\revised{\today}

\begin{document}

\title{TDEs on FIRE: Illuminating the Cosmic Evolution of Tidal Disruption Rates}

\author[orcid=0000-0003-2694-933X,sname='Kar Chowdhury', gname='Rudrani']{Rudrani Kar Chowdhury}
\affiliation{Department of Physics, The University of Hong Kong, Pokfulam Road, Hong Kong, China}
\affiliation{Tata Institute of Fundamental Research, Homi Bhabha Road, Mumbai 400005, India}
\email[show]{rudrani.chowdhury@tifr.res.in}  

\author[orcid=0000-0002-9589-5235, gname=Lixin,sname=Dai]{Lixin Dai}
\affiliation{Department of Physics, The University of Hong Kong, Pokfulam Road, Hong Kong, China}
\affiliation{The Hong Kong Institute for Astronomy and Astrophysics, The University of Hong Kong, Pokfulam Road, Hong Kong, China}
\email[show]{lixindai@hku.hk}

\author[orcid=0009-0004-2575-1924,gname=Janet N.Y., sname='Chang']{Janet N.Y. Chang} 
\affiliation{Department of Physics, The University of Hong Kong, Pokfulam Road, Hong Kong, China}
\affiliation{The Hong Kong Institute for Astronomy and Astrophysics, The University of Hong Kong, Pokfulam Road, Hong Kong, China}
\email{janetcny@connect.hku.hk}

\author[0000-0003-2544-054X,sname=Tsang Keung,gname=Chan]{Tsang Keung Chan}
\affiliation{Department of Physics, The Chinese University of Hong Kong, Shatin, Hong Kong, China}
\email{tsangkeungchan@cuhk.edu.hk}
 
\begin{abstract}

\noindent

Tidal disruption events have been extensively studied in the local Universe, but their prevalence at high redshifts remains largely unexplored. Using the FIRE-2 cosmological zoom-in simulations, we compute the per-galaxy tidal disruption rate (TDR) over $z = 1$–$10$, covering black holes from IMBHs to SMBHs. The averaged TDR rises from the early Universe, peaks at $\sim 4 \times 10^{-4} \, \text{gal}^{-1} \text{yr}^{-1}$ near $z \sim 2.5$, and declines to $\sim 10^{-5} \, \text{gal}^{-1} \text{yr}^{-1}$ at $z=1$. 
The TDR correlates strongly with host galaxy star formation rate and central stellar density at all redshifts. Qualitatively, the TDR trends with the $M_{\rm BH}$ and $M_{\rm gal}$ persist from high redshift to the local Universe, suggesting similar black hole-galaxy scaling across cosmic time. Satellite galaxies exhibit comparably high TDRs, with their fractional contribution increasing significantly at high redshifts, highlighting their potential for probing IMBHs and early galaxy assembly. This work demonstrates that cosmological simulations offer a promising avenue for constraining the cosmic evolution of the TDR, paving the way for future comparisons with next-generation observations.

\end{abstract}

\section{Introduction} \label{sec:intro}

Two-body scattering at a galactic center can send a star onto an orbit that brings it too close to a massive black hole (MBH) \citep{Magorrian99}. If the star passes within the tidal radius of an MBH, tidal forces tear it apart. Roughly half of the debris falls back onto the MBH, producing a luminous flare that peaks in weeks and fades over months \citep{Rees88, Evans89}. This type of transient is called a tidal disruption event (TDE). The properties of a TDE reveal critical information about both the MBH and the disrupted star \citep{Guillochon13, Leloudas16, Mockler19}. Recent studies have further shown that TDEs are powerful probes of extreme accretion and outflow physics around MBHs \citep{Dai21}, intermediate-mass black holes \citep[IMBHs;][]{Chang24}, and the first stars in the Universe \citep{rkc24}. In addition, statistical TDE samples and event rates reveal large-scale host galaxy properties \citep{Stone16, French20, Pfister20, Chang26}.

The redshift evolution of TDE rates, however, remains almost entirely unconstrained, both observationally and theoretically (however, see \cite{Kochanek16} and \cite{Chen20} for volumetric TDE rates across $z \simeq 1-6$). While observing TDEs at $z = 1-10$ would reveal black hole growth, demographics, and coevolution with host galaxies, the current TDE sample is overwhelmingly local because surveys like the Zwicky Transient Facility are flux limited. Only a handful of jetted TDE candidates exist at $z\sim1$ and their detection relies on large beamed luminosities \citep{Cenko12, Brown15, Andreoni22}.
Theoretical estimates face even deeper uncertainties. The TDE rate prediction sensitively depends on the galaxy stellar profiles, which remain highly uncertain for high-redshift galaxies. Even recent JWST observations \citep{Cochrane25} have not resolved tensions with the theory \citep{Labbe23}. The TDE rate is also highly sensitive to the black hole mass function (BHMF), yet low-mass MBHs, which dominate the TDE population, are poorly constrained at high redshift \citep{Kelly12, Taylor25}. As a result, the cosmic evolution of TDE rates remains without any robust prediction or measurement.

A recent breakthrough offers a glimpse forward: \cite{Karmen25} detected a TDE candidate at $z\sim5$ in the JWST COSMOS-Web survey \citep{Casey23}. This discovery opens the door to finding more early Universe TDEs. The number of such events is expected to grow substantially with high-cadence, wide-area facilities like eROSITA \citep{Sazonov21}, the Einstein Probe \citep{yuan25}, the Vera Rubin Observatory \citep{Ivezic19}, and the Nancy Grace Roman Space Telescope \citep{spergel15}, and with future mission such as ULTRASAT \citep{Sagiv14}. These same missions will open the door to novel TDE classes, such as off-nuclear (ON) TDEs from IMBHs in dwarf galaxies and globular clusters (GCs), and TDEs from SMBH binaries \citep{Lin18, Angus22, Jin25, yao25, Patra26}, each offering new windows into galaxy assembly and merger histories \citep{Ricarte21}.

It is therefore timely to investigate the TDE rates and their dependence on host galaxy properties across cosmic time, spanning black holes from the centers of massive galaxies to the small satellite scales. Previous zoom-in simulations studied TDEs at very high redshifts ($z > 6$) \citep{Pfister19, Pfister21, Lee23}, but only for small galaxy samples and at redshifts beyond the detectability of most current surveys. 
The second generation of the Feedback In Realistic Environments (FIRE-2) simulation \citep{Wetzel23} is ideally suited for this study, offering superior resolution and comprehensive modeling of star formation, evolution, and stellar feedback compared to other existing simulations.
Our goal is to cover the redshift range $z=1-10$ with better-resolved stellar structures and a statistically larger sample than previous work, and to use this framework to predict TDE rates across the massive primary galaxies to smaller satellite galaxies and their detectability with current and upcoming surveys.

In this paper, we have used the Massive Halo simulation suite of FIRE-2 available across a redshift range of $z=1-10$. The novelty of the Massive Halo suite lies in the inclusion of the sophisticated model of the formation and growth of MBHs in the simulation \citep{Angles17}, which is crucial for the purpose of this paper. Additionally, this simulation suite offers the unprecedented spatial resolution of FIRE-2 that reaches a few parsecs scale, and careful and comprehensive modeling of star formation, along with the evolution and stellar feedback physics. These are discussed in Section \ref{sec:simulation}, following the introduction of the framework for theoretical estimation of TDE rates in Section \ref{sec:lc}. We describe our methodology of combining the two and obtaining the TDE rate in Section \ref{subsec:sim_rate}. Afterwards, we describe the findings from this study in Section \ref{sec:result}, including the redshift evolution of TDE rates (Section \ref{sec:z_evol}), their correlation with the star formation rate (Section \ref{sec:SFR-TDR}), host galaxy and MBH masses (Section \ref{sec:mass_scaling}), and the stellar properties (Section \ref{sec:TDR-host props}). TDE rates in the satellite galaxies are discussed in Section \ref{sec:ON}. Finally, in Section \ref{sec:summary}, we summarize the results, discuss future research directions based on the foundation of this paper, and certain limitations of the current work.

\section{Methodology} \label{sec:theory}

{\noindent}We provide a general background of calculating the TDE rates assuming two-body scattering between the stars in the following section. A basic overview of the TDE rate calculation using loss cone theory is discussed in Section \ref{sec:lc}. We introduce FIRE-2 simulation in \ref{sec:simulation}, with a brief discussion of the important parameters of the Massive Halo suite and black hole model given in Sections \ref{subsec:sim_par} and \ref{subsec:bh}, respectively. Finally, calculation of the intrinsic TDE rates in the simulated FIRE-2 galaxies is discussed in Section \ref{subsec:sim_rate}.

\subsection{TDE Rate from Loss Cone Theory} \label{sec:lc}
{\noindent}TDE rates are calculated based on the loss cone dynamics, which we briefly describe here and refer to \cite{Stone20} for a more complete discussion. The loss cone can be understood as a region in phase space through which stars diffuse into orbits close to the MBH and become disrupted. The distance to the MBH at which stars get disrupted is called the tidal radius and can be approximated as 
\begin{equation}
    R_T = r_\star\left(\frac{M_{\rm BH}}{m_\star}\right)^{1/3},
\end{equation}
where $M_{\rm BH}$ is the mass of the black hole (BH), and $m_\star$ and $r_\star$ are the mass and radius of the disrupted star, respectively. It is common practice to define a variable, called the penetration parameter ($\beta$), such that
\begin{equation}
    \beta = \frac{R_T}{R_p},
\end{equation}
where $R_p$ is the pericenter of the stellar orbit.
Correspondingly, the specific angular momentum of the stellar orbit at $R_T$ can be written as
\begin{equation}
    L_{\rm lc} = \sqrt{2GM_{\rm BH}R_T}~,
\end{equation}
Hence, stars with $L<L_{\rm lc}$ will get disrupted within $R_T$. The phase space corresponding to $L<L_{\rm lc}$ is called the loss cone.

The number of stars per unit specific binding energy $(E)$ and $\beta$ scattered into the loss cone due to collisional two-body relaxation is estimated as \citep{Strubbe11, Pfister20, Chang24}:
\begin{equation}
     \frac{d^2\Gamma}{dE~d\ln \beta} = 8 \pi^2 G M_{\rm BH} \frac{R_T}{\beta} \mathcal{F}(q(E), \beta),
     \label{eq:flux}
\end{equation}
where 
\begin{align}
\begin{split}
    \mathcal{F}(q(E),\beta) ={} & \frac{f(E)}{1+q^{-1}\xi(q) \ln (1/R_{\text{lc}})}\\
    & \bigg[1-2 \sum^{\infty}_{m=1} \frac{e^{-a^2_m q/4}}{a_m} \frac{J_0(a_m \beta^{-1/2})}{J_1(a_m)}\bigg],
    \label{eq:rate:comp1}
\end{split}\\
\xi(q)={} &  1-4\sum^\infty_{m=1} \frac{e^{-a^2_m q/4}}{a_m^2}.
\label{eq:rate:comp2}
\end{align}
Hence, the total consumption rate of stars into the loss cone gives the TDE rate
\begin{equation}
    \Gamma = \int \int \frac{d^2\Gamma}{dE~d\ln \beta} dE~d\ln \beta~.
    \label{eq:TDR}
\end{equation}

Different terms in Equations \ref{eq:rate:comp1} and \ref{eq:rate:comp2} are as follows:
\begin{itemize}
    \item $f(E)$: Stellar distribution function, which is the number density of stars with specific energy $E$ in the phase space. This can be obtained from the density profile of the stars ($\rho$) and total gravitational potential ($\psi$) through Eddington's formula \citep{BT87} 
    \begin{equation}
    f(E)= \frac{1}{\pi^2m_\star\sqrt{8}}\frac{d}{dE}\int_0^E \frac{d\rho}{d\psi} \frac{d \psi}{\sqrt{E-\psi}}~.
    \label{eq:DF}
\end{equation}
    \item $q(E)$: Loss cone filling factor that demarcates between the full ($q \gg 1$) and empty ($q \ll 1$) loss cone regimes and is given by \begin{equation}
    q(E)=\frac{P(E)\overline{\mu}(E)}{R_{\rm lc}(E)}~,
    \label{eq:q}
\end{equation}
where $\overline{\mu}(E)$ is the orbit-averaged angular momentum diffusion coefficient, $P(E)$ is the orbital period of stars, and \begin{equation}
    R_{\rm lc}(E) = L_{\rm lc}^2/L_c^2(E),
    \label{eq:Rlc}
\end{equation}
$L_{c}$ being the angular momentum of a circular orbit.
    \item $J_0,J_1$: The Bessel functions of zeroth and first order, respectively. $a_m$ is the $m ^{\rm th}$ zero of $J_0$.
\end{itemize}

The differential TDE rate (Equation \ref{eq:flux}) reaches a maximum near a critical energy, $E_{\rm crit}$, producing the highest flux into the loss cone at the energy when $q(E_{\rm crit})=1$. This energy corresponds to a critical distance from the BH, denoted as $r_{\rm crit}$ \citep{Stone16}. For BHs with mass $< 10^8M_\odot$, this scale coincides with the radius of influence, $r_{\rm inf}$, defined as the radius at which total enclosed stellar mass is equal to the mass of the BH \citep{Stone16} (Appendix \ref{app:rate_eq}). Hence, resolving the stellar density profile at $r_{\rm inf}$ becomes critical for a proper estimation of TDE rate in the loss cone framework. In the following sections, we provide a description of the FIRE-2 simulation and TDE rates calculated using stellar distributions in FIRE-2 galaxies.

\subsection{FIRE-2 Simulation} \label{sec:simulation}

{\noindent} We briefly summarize below the components of FIRE-2 relevant for this paper and refer to \cite{Wetzel23} for a more detailed description of the simulation suites and discussion of their results.

\subsubsection{FIRE-2 Massive Halo Suite Parameters} \label{subsec:sim_par}

{\noindent}In this work we use four simulations; A1, A2, A4, and A8, collectively known as the Massive Halo suite in the public data release of FIRE-2.\footnote[1]{We note that the second data release of FIRE-2 contains more galaxies, and a larger redshift range with more snapshots \citep{Wetzel25}. However, the entire analysis of the paper was completed before the public release of DR2 FIRE-2 and we plan to use this bigger sample in future studies.} The Massive Halo suite is resimulated from the MassiveFIRE galaxy simulation evolved with the FIRE-2 code \citep{Feldmann17}. FIRE-2 is a cosmological zoom-in simulation, evolved with the $N$-body$+$hydrodynamics code GIZMO \citep{Hopkins15}. The simulations are developed to model galaxy formation in the cosmological simulation framework aiming to resolve multiphase interstellar medium by implementing realistic stellar evolution and feedback mechanisms. FIRE-2 includes radiative cooling, heating, low-energy ionizing cosmic rays, dust, star formation, and stellar feedback in terms of stellar wind, core collapse, and Type Ia supernovae. We refer to \cite{Hopkins18} for a detailed description of the simulation models and associated parameters of FIRE-2.

All four simulations in Massive Halo can achieve the mass resolution of $m_b = 3.3 \times 10^4 M_\odot$ and $m_{\rm DM} = 1.7 \times 10^5 M_\odot$ for the baryonic and dark matter (DM) particles, respectively. The force softening length of DM, black hole, star, and gas particles are $\epsilon_{\rm DM}=57$ pc, $\epsilon_{\rm BH}=7$ pc, $\epsilon_{\rm star}=7$ pc, and $\epsilon_{\rm gas}=0.7$ pc, respectively. Adaptive force softening is used for gas particles with a minimum length of $\epsilon_{\rm gas}$, whereas $\epsilon_{\rm DM}, \epsilon_{\rm BH}$ and $\epsilon_{\rm star}$ are kept fixed below $z=9$. Such a high resolution achieved in FIRE-2 makes it exceptional among the existing cosmological simulations. This outstanding resolution is crucial for determining the stellar distribution at the galaxy centers to accurately estimate their TDE rates. Total stellar mass at $z=1$ reaches $2.75\times 10^{11}M_\odot, 4.10\times 10^{11}M_\odot, 2.34\times 10^{11}M_\odot$, and $5.36 \times 10^{11}M_\odot$ in A1, A2, A4, and A8, respectively. We briefly discuss below the seeding criteria and evolution of black holes in the Massive Halo and refer the reader to \cite{Angles17} for detailed discussions.

\subsubsection{Formation and Growth of Black Holes in FIRE-2}\label{subsec:bh}
\noindent
Black holes are considered to be collisionless sink particles. Seed BHs of mass $M_{\rm seed} =1.4 \times 10^4 M_\odot$ are formed using the friends of friends algorithm in the DM halo by converting the most-bound star particle when the stellar mass of a DM halo exceeds $1000\times M_{\rm seed}$ \citep{DiMatteo08}. Such seed BHs grow afterwards through accretion and merger \citep{Springel05}. We note that feedback from the BH is not incorporated in the Massive Halo suite, although the accretion rate is adjusted to match the normalization of the observed $M_{\rm BH}-M_{\rm bulge}$ relation in the local Universe. The impact of the absence of active galactic nuclei (AGN) feedback on the results of this paper is discussed in Section \ref{sec:summary}.

Furthermore, the resolution of the cosmological simulation limits the ability to fully resolve the gravitational dynamics of the black holes. While in reality, BHs have much higher mass than the surrounding gas, dark matter, and individual star particles, this might not always be the case in the simulation due to numerical artifacts. A seed BH could sometimes form with a lower mass compared to the other particles, which might prevent it from sinking to the gravitational potential of the galaxies or even cause the BH to be removed from the galaxy in the absence of dynamical friction. To prevent this, individual black holes are given an artificial dynamical mass of $m_{\rm BH}=300 \times m_b$ which is independent of their physical mass $M_{\rm BH}$ gained through accretion. However, if no MBH is present at the center of the galaxy due to dynamical friction at a particular time step, we consider the nearest black hole to the host galaxy center as the central black hole throughout this paper.

\subsection{TDE Rate Calculations from Simulated Galaxies \label{subsec:sim_rate}}

\begin{figure*}
\centering
    \includegraphics[width=17cm]{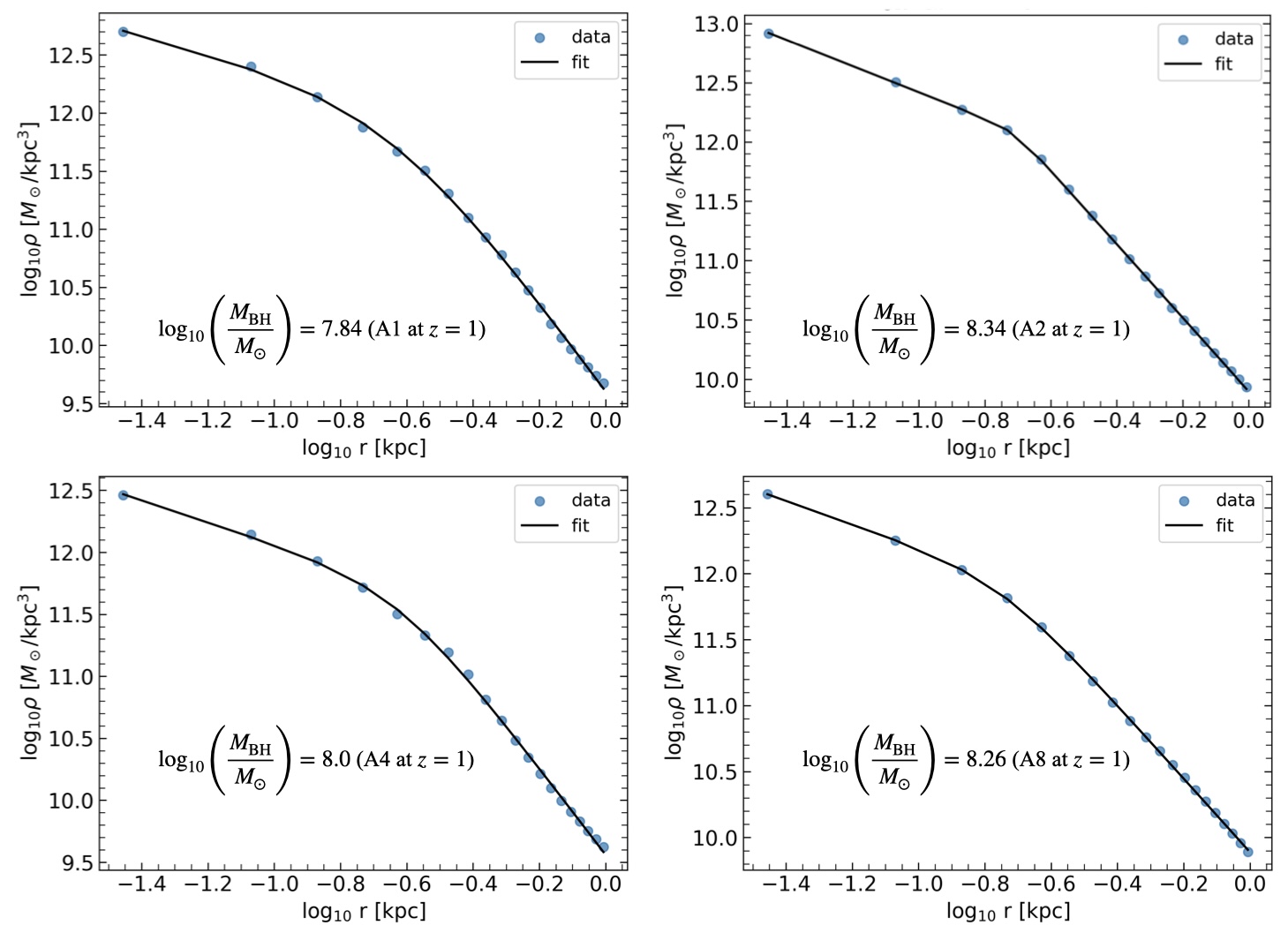}
    \caption{The radial profile of stellar density at the centers of the primary host galaxies at $z=1$ in the A1, A2, A4, and A8 runs, respectively. Double power-law fits (Equation \ref{eq:dp_profile}) are shown as solid black lines. The primary galaxies of A2 and A8 do not contribute to the TDE rates as their BH masses are $>10^8M_\odot$.}
    \label{fig:stellar_density}
\end{figure*}

\noindent To compute the TDE rates of the simulated galaxies, we mainly focus on the properties and dynamics of the stars in the nuclear regions of the galaxies across the redshift range $z=1-10$. We begin by selecting the galaxies with total stellar mass $M_{\rm gal} > 10^7 M_\odot$ in the four simulation runs A1, A2, A4, and A8, which is the threshold halo mass above which seed BHs are formed in FIRE-2, as discussed in Section \ref{subsec:bh}.

We then obtain the stellar densities around these individual galaxies. Examples of stellar distribution and radial stellar density profiles at the galaxy centers are shown in Appendix \ref{app:rate_eq}. Figure \ref{fig:maps} shows the individual projected stellar densities at the centers of the four primary host galaxies at $z=1$ for A1, A2, A4, and A8 runs, respectively. Using these stellar densities, we first obtain the radial density profiles of stars at the center of each galaxy within $1$ kpc radius. The reason for choosing the galaxy center instead of the central BH position lies in the occasional absence of BHs at the galaxy center due to dynamical friction as discussed above. Furthermore, we fit each stellar density profile with a double power-law function as follows:
\begin{align}
        \rho (r)= \rho_0\left(\frac{r}{r_0}\right)^{- \gamma}\left[1+\left(\frac{r}{r_0}\right)^\alpha\right]^{\frac{\gamma-\delta}{\alpha}},
    \label{eq:dp_profile}
\end{align} 
where $\rho_0$ is the volume density of stars at scale radius $r_0$. $\gamma$ and $\delta$ are the power-law indices for the inner and outer density profiles, respectively, and $\alpha$ denotes the transition between the slopes of these two power-law profiles.

It is worth noting that Equation \eqref{eq:dp_profile} assumes spherical symmetry for the stellar distribution, and the same assumption underlies the TDE rate calculation within the loss cone framework. While real galaxies -- whether observed or simulated -- are rarely perfectly spherical, most previous studies that estimate TDE rates from observed galactic profiles have nonetheless adopted this simplification \citep[e.g.,][]{Stone16, Pfister20, Pfister21, Chang24}, owing to the absence of an analytical framework for computing TDE rates in asymmetric or triaxial stellar systems. Therefore we also adopt this approach in this work.

However, we recognize that the stellar distributions in some systems from the Massive Halo suite, particularly among smaller galaxies, do not even appear spherically symmetric. To mitigate potential biases, we visually inspected all galaxies across the four Massive Halo runs and divided the sample into two categories: those that visually appear spherically symmetric, and those that show significant deviations from spherical symmetry. While we restrict our main analysis to the spherically symmetric subsample, it is worth considering how our exclusion of asymmetric systems might affect the overall results. Theoretically, triaxiality or asymmetry could either enhance or suppress TDE rates relative to the spherical case, depending on the detailed orbital structure and the efficiency of angular momentum diffusion. A definitive assessment of the net effect would require a full nonspherical loss cone framework, which is beyond the scope of this paper. We therefore refrain from speculating on the direction of any potential bias, and instead emphasize that our conclusions are robustly derived for the spherically symmetric subsample, with the comparison between both samples provided in Appendix \ref{app:asymmetric}.

Based on the fitted parameters of the stellar density profiles (Equation \ref{eq:dp_profile}), we further constrain our galaxy sample. We discard the nonphysical systems with inner slopes $(\gamma)$ steeper than their outer slopes $(\delta)$ and select only those with $\gamma < \delta$ \citep{Hopkins20, Horta24}. A large spurious inner slope could be attributed to the numerical noise of the simulation caused by a handful of star particles present in the galaxy centers. An example of such stellar profile and the corresponding stellar distribution map are shown in Figure \ref{fig:app_stellar_density}. Moreover, we restrict our sample to those with $0 < \gamma <3$ due to numerical reasons. The lower limit on $\gamma$ ensures a positive stellar distribution function (Equation \ref{eq:DF}), whereas the upper limit prevents the enclosed stellar mass from diverging. Furthermore, we discarded the BHs with $M_{\rm BH} > 10^8M_{\odot}$ beyond which no TDEs are produced as the stars are captured as a whole by the BHs. Figure \ref{fig:stellar_density} shows examples of stellar profiles at the centers of the primary host galaxies at $z=1$ in the A1, A2, A4, and A8 runs. It should be noted that $M_{\rm BH} > 10^8M_{\odot}$ in the A2 and A8 runs. Hence, these two systems do not contribute to the TDE rates at $z=1$ in the subsequent analysis. 

We estimate the primary parameters for calculating the rates of TDEs for this conservative galaxy sample using the publicly available code \textsc{phaseflow} \footnote[2]{\textsc{phaseflow} is included in the \textsc{agama} software library \\ \href{https://github.com/GalacticDynamics-Oxford/Agama}{https://github.com/GalacticDynamics-Oxford/Agama}} \citep{Vasiliev17, Vasiliev19}. The \textsc{phaseflow} code is used to solve for the distribution function $f(E)$ from the density profile by Eddington inversion (Equation \ref{eq:DF}) using the fitted parameters of the double power-law profile ($\rho_0$, $r_0$, $\alpha$, $\delta$ and $\gamma$). Furthermore, it calculates the combined potential of the stars and central BH by solving the Poisson equation \citep{Vasiliev17, Vasiliev19}. \textsc{phaseflow} can also solve for the circular angular momentum $(L_c)$, orbit-averaged angular momentum ($\overline{\mu}$), orbital period of stars $P(E)$, and loss cone filling factor $q$ (Equation \ref{eq:q}). With this information, we calculate the TDE rates per galaxy per unit time using Equation \ref{eq:TDR}. Throughout this paper, we have assumed a monochromatic stellar population, where all the stars are Sun-like with $m_\star =1M_\odot$ and $r_\star=1R_\odot$. In the following sections we present the TDE rates across $z=1-10$ and their correlation with MBH and host galaxy stellar properties. 

\begin{figure*}
    \centering
    \includegraphics[width=18cm]{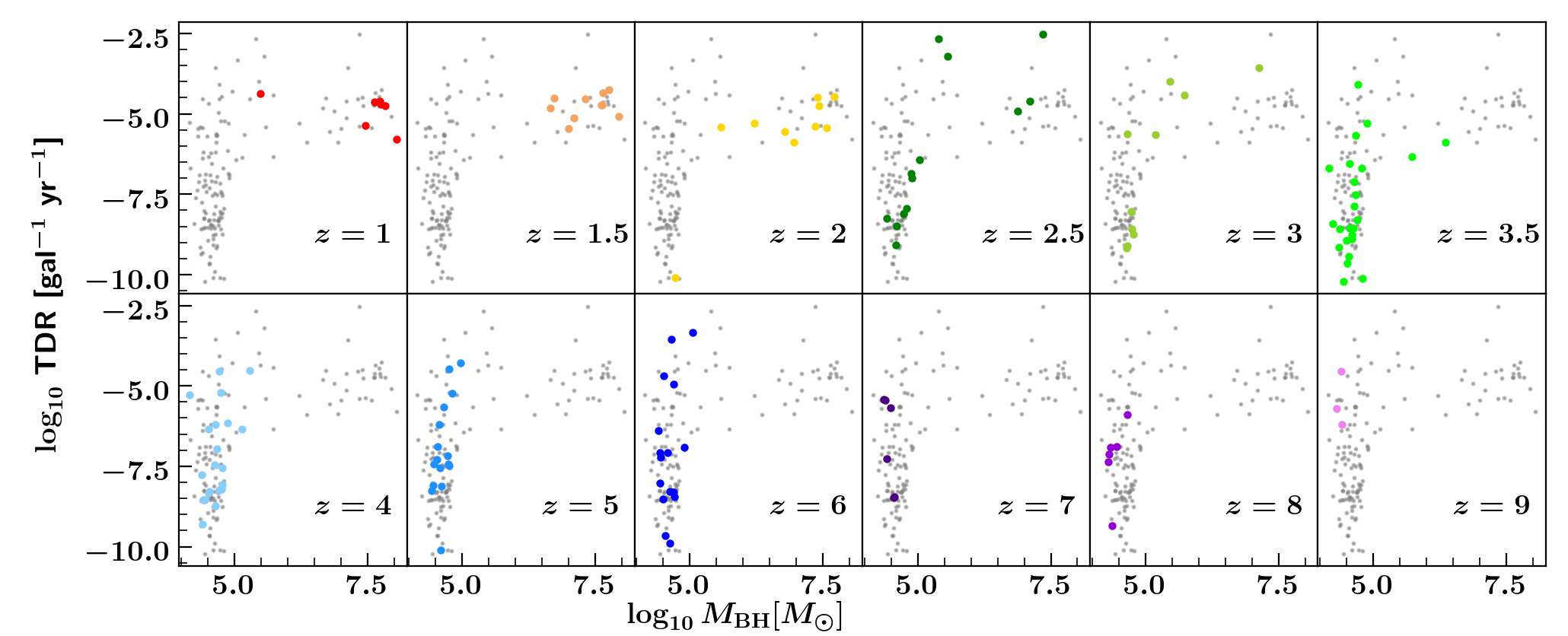}\\
    \caption{The correlation between the TDR and $M_{\rm BH}$ at individual redshifts. Gray dots show the full galaxy sample across all redshifts, while colored points in each panel highlight galaxies at the specific redshift indicated.}
    \label{fig:scaling1_z}
\end{figure*}
\begin{figure*}
    \centering
    \includegraphics[width=18cm]{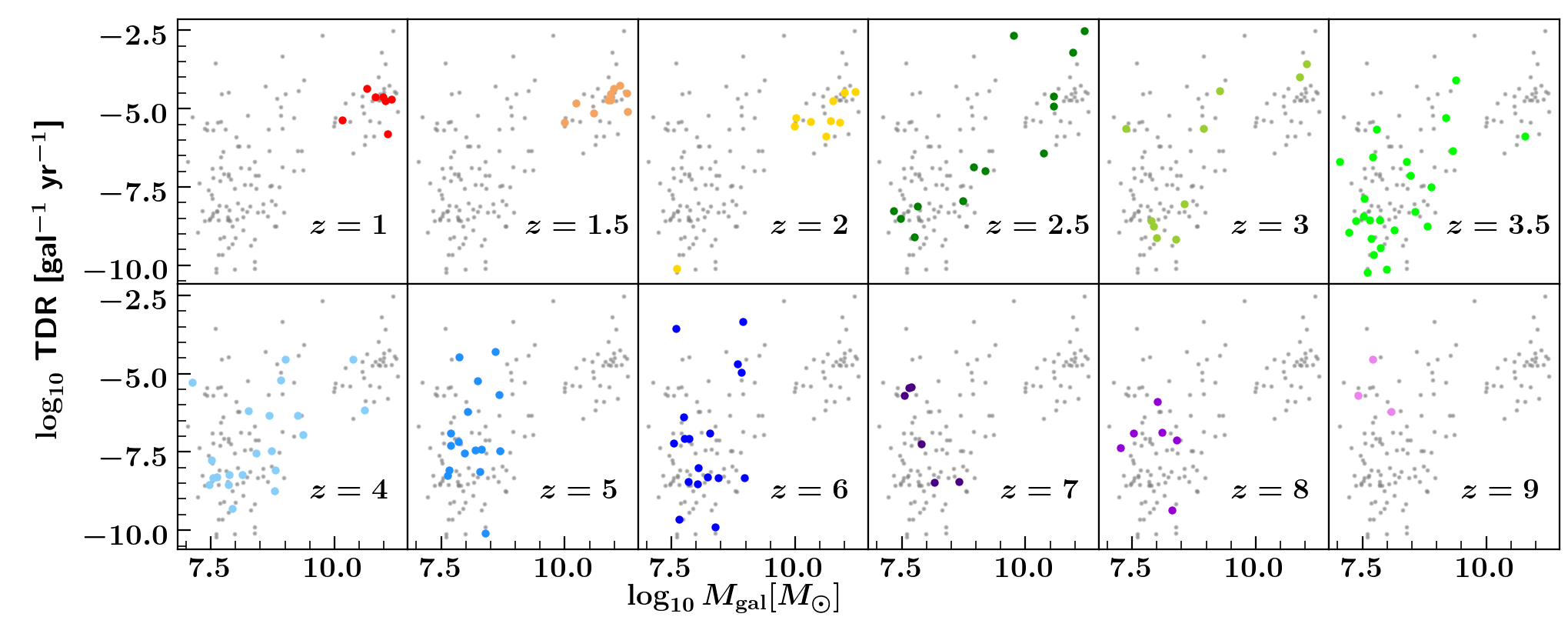}
    \caption{Correlation between the TDR and $M_{\rm gal}$ at individual redshifts. Color scheme is same as Figure \ref{fig:scaling1_z}.}
    \label{fig:scaling2_z}
\end{figure*}

\section{Results} \label{sec:result}

\noindent We study the TDE rates (TDR) in FIRE-2 galaxies and the corresponding redshift evolution in Section \ref{sec:z_evol}, followed by their correlation with the star formation rate in Section \ref{sec:SFR-TDR}. Correlated properties of the TDR with their host galaxy and BH masses are studied in Section \ref{sec:mass_scaling}. Further investigation of the impact of the host galaxy stellar properties on the TDR is done in Section \ref{sec:TDR-host props}. Finally, we examine the TDR in the satellite galaxies and their detectability with the latest telescopes in Section \ref{sec:ON}.

\subsection{Redshift Evolution of TDE Rates}\label{sec:z_evol}

\noindent We begin by calculating the TDR for the individual galaxies at specific redshifts, and show these rates as a function of the black hole mass (Figure~\ref{fig:scaling1_z}) and the galaxy mass (Figure~\ref{fig:scaling2_z}).  
The galaxy sample, sparse at $z \gtrsim 7$, grows over time. Small galaxies and IMBHs dominate at $4 \lesssim z \lesssim 6$; massive galaxies (and SMBHs) appears by $z \sim 3.5$, which gradually merge to become very massive galaxies by $z \lesssim 2.5$. 
Despite the large scatter in the TDRs at fixed redshift, the figures hint at an increasing trend with the $M_{\rm BH}$ and $M_{\rm gal}$ in the early Universe.

\begin{figure}
    \includegraphics[width=1\linewidth]{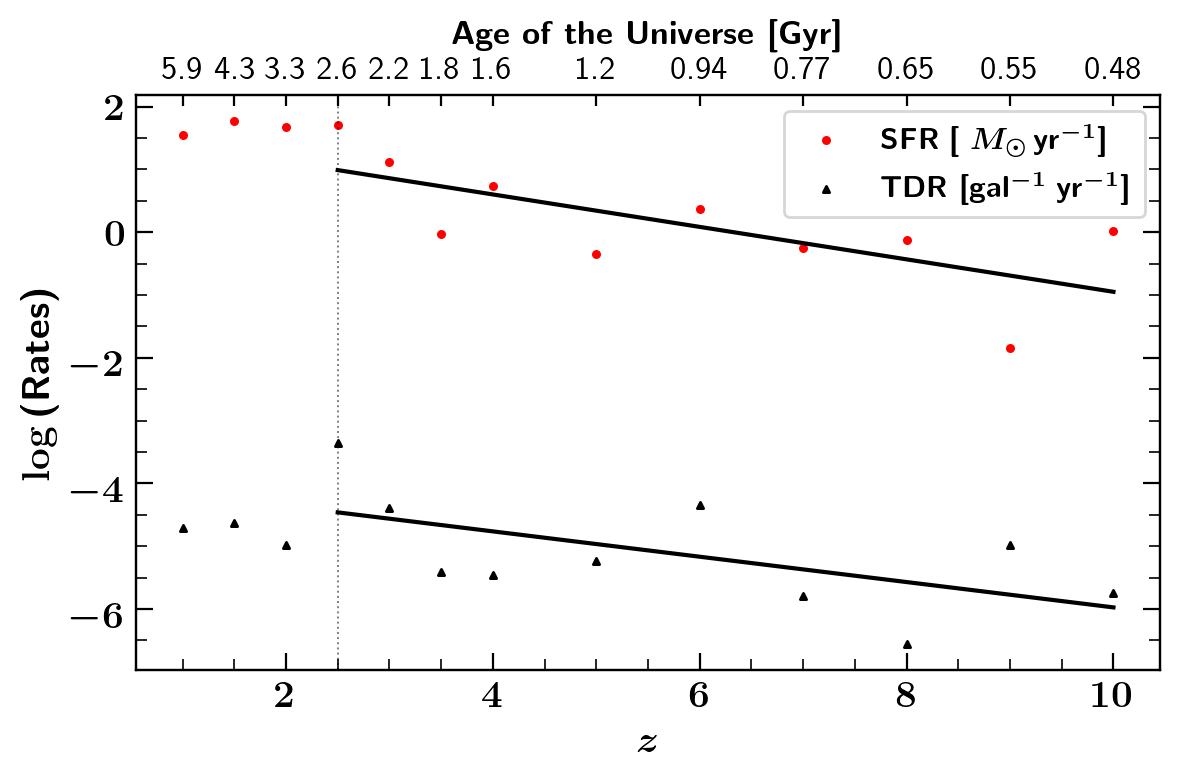}
    \caption{The redshift evolution of the averaged TDR (black triangles) and the averaged SFR (red dots) at each redshift. Black solid lines show the best-fit relations at $z > 2.5$, while the thin dotted line marks the redshift below which both rates moderately decline.}
    \label{fig:ratez}
\end{figure}

To understand how the TDR evolves across the cosmic time, we plot in Figure~\ref{fig:ratez} the averaged TDR at individual redshifts as a function of $z$. TDR appears to peak around $z \sim 2-3$, consistent with the Figures \ref{fig:scaling1_z} and \ref{fig:scaling2_z}. We fit the high-redshift TDR using a power-law function of $z$, and obtain the following best-fit relation:
\begin{equation}
    \log_{10} \frac{\rm TDR}{\rm gal^{-1} yr^{-1}} = z^{-0.2} - 4;~z\ge2.5.
    \label{eq:ratez}
\end{equation}
We further notice a moderately decreasing TDR at $z<2$, although caution is needed when interpreting this trend, as the simulated galaxies were evolved only down to $z=1$. However, we understand from the cosmic star formation history that the SFR declines at low redshift \citep{Madau14}. Hence, we expect a decline in the TDR in the local Universe if the TDR follows the SFR as noted at high redshift.

Such evolution of the TDR with redshift could be attributed to the black hole model, star formation rate, and their host galaxy stellar properties in FIRE-2, which we investigate in detail in the following sections. 
The redshift evolution of the SFR is also shown in the same plot for a comparison, which will be further discussed in the next section. 

\subsection{Impact of Star Formation Rate on TDR} \label{sec:SFR-TDR}

\begin{figure*}
    \centering
    \includegraphics[width=15cm]{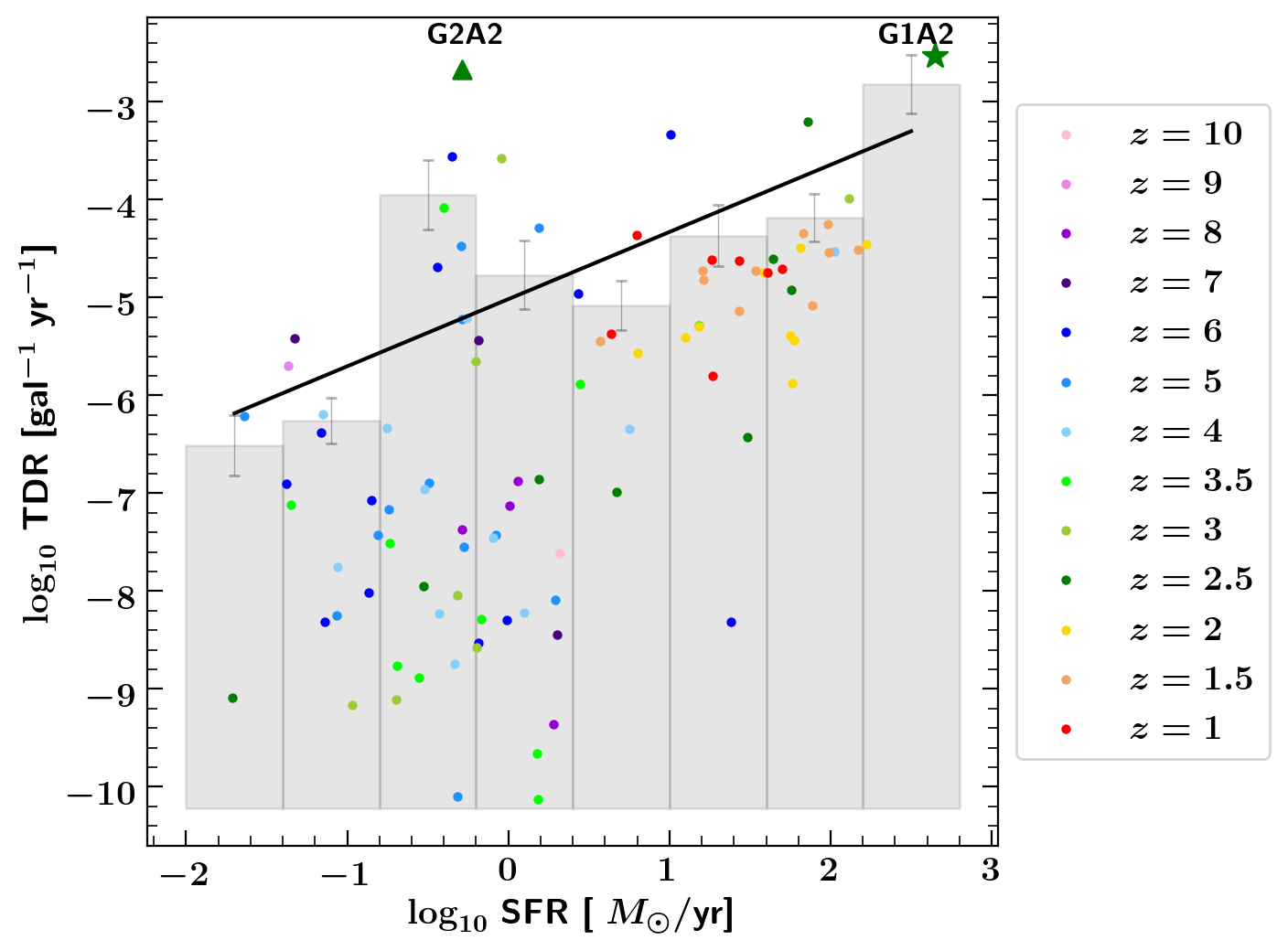}
    \caption{The TDR versus SFR for individual galaxies in our sample, with the colors indicating redshift from $z=1$ to $10$. The TDR generally traces the redshift evolution of SFR. The gray-shaded regions show the averaged TDR in each SFR bin, fitted with a power-law function  (Equation~\ref{eq:TDR-SFR}).  }
    \label{fig:sfr}
\end{figure*}

\noindent
In this section, we investigate how the TDR and the star formation rates (SFRs) are correlated in the host galaxies. In FIRE-2, stars are formed from the dense molecular self-gravitating gas at a rate $\dot{\rho}_\star=\rho_{\rm gas}/t_{\rm ff}$, where $t_{\rm ff}$ is the freefall timescale. Instantaneous SFR is stored in the simulation. We refer the reader to \cite{Hopkins18} for the details on star formation and stellar feedback implemented in FIRE-2.
For the simulated galaxies, we calculate the SFR within a central radius that contains $90 \%$ of the stellar mass of the host galaxy. As noted from the Figure \ref{fig:ratez}, TDR closely follows SFR, where both increase as redshift decreases from $z=10$ until $z\sim 2-3$, below which the SFR appears to get saturated. However, the comparison of the TDR with the SFR is based on the particular snapfiles at discrete redshifts. Hence, a direct comparison of these two physical quantities might not be straightforward. Nevertheless, we notice, at least at high redshift, the peak of TDR is following the peak of SFR. This might be indicative of some other mechanism, following a starburst phase, that is contributing to the enhanced TDR. This hints toward a similar phenomenon in the local Universe where an abundance of TDEs is found in the post-starburst galaxies \citep{French16, Graur18}.

We further investigate in Figure \ref{fig:sfr} the correlation between the TDR and the SFR in individual galaxies, along with their fitted scaling relation. Average TDR is calculated in each SFR bin along with the corresponding standard error on the mean. We notice that the averaged TDR is well correlated with the SFR across the entire redshift range of $z=1-10$. TDR can be fitted as a simple power-law function of the SFR as follows:
\begin{equation}
    \rm{TDR} = 9.6 \times 10^{-6} \left(\frac{\rm{SFR}}{M_\odot \rm{yr}^{-1}}\right)^{0.69} \rm{gal}^{-1}~yr^{-1}.
    \label{eq:TDR-SFR}
\end{equation}

Two galaxies in Figure~\ref{fig:sfr} produce very high TDRs ($>10^{-3} \, \text{gal}^{-1} \, \text{yr}^{-1}$). Both are located at $z=2.5$ in the A2 run, labeled `G1A2' (dark green star, the primary host) and `G2A2' (dark green triangle). Despite both having the highest TDRs, G2A2 has a much lower SFR ($\sim 0.5 \, M_{\odot} \, \text{yr}^{-1}$), while G1A2 shows the highest SFR among all the Massive Halo simulations ($\sim 450 \, M_{\odot} \, \text{yr}^{-1}$). Therefore, it is clear that additional factors beyond the SFR must influence the TDR of host galaxies, which motivated us to investigate how the host galaxy stellar properties impact the TDR. We will discuss this in Section \ref{sec:TDR-host props}.

\subsection{Dependence of TDE Rates on  $M_{\rm gal}$ and $M_{\rm BH}$}\label{sec:mass_scaling}
\begin{figure*}
    \centering
    \includegraphics[width=18cm]{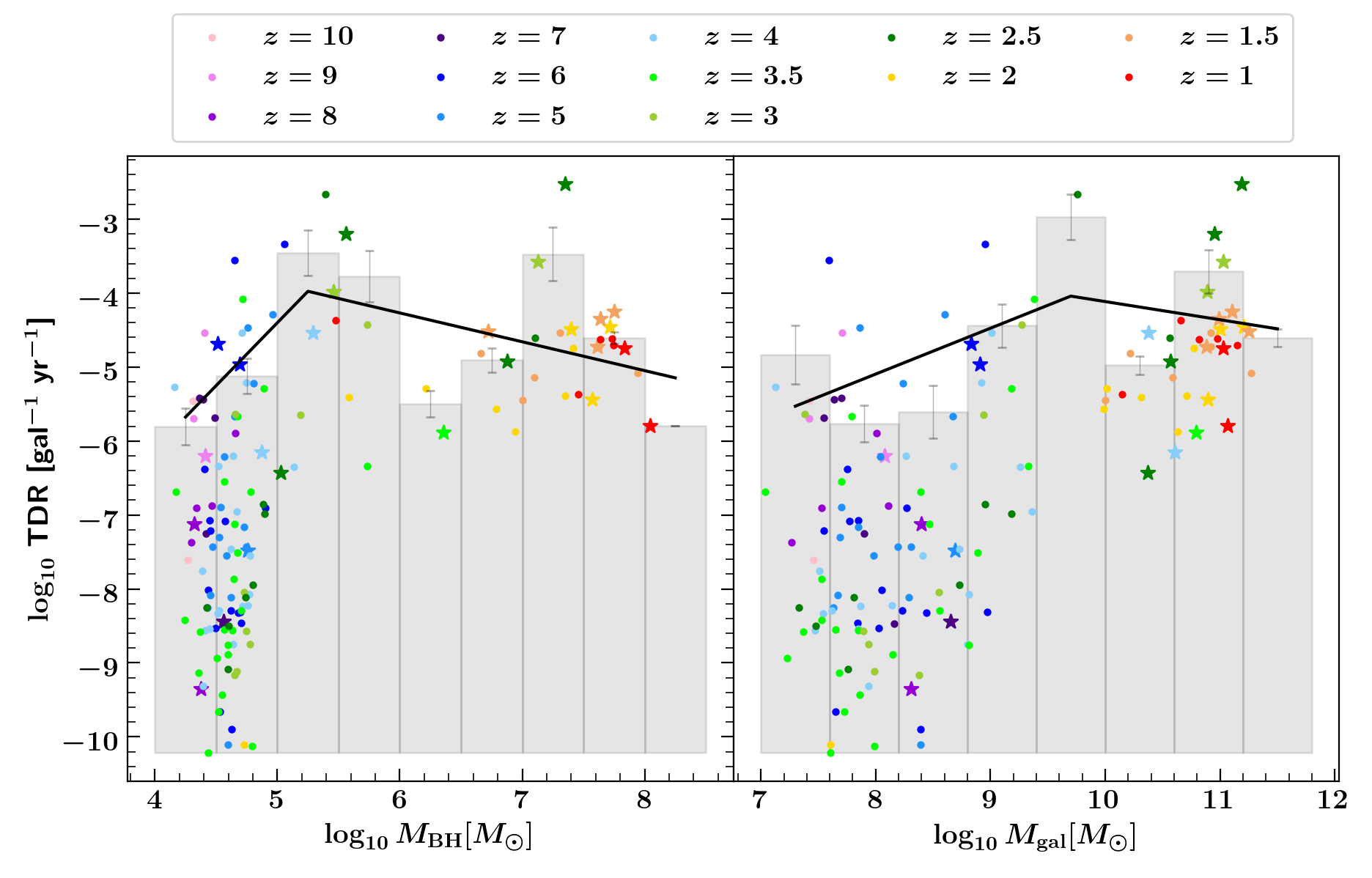}
    \caption{The correlation of the TDR with the black hole mass $(M_{\rm BH})$ (left panel) and host galaxy mass $(M_{\rm gal})$ (right panel) for individual galaxies in our sample. Colors indicate different redshifts, as shown in the legend. In both panels,  star symbols denote primary host galaxies, while circles represent the remaining smaller galaxies. The gray-shaded regions indicate the average TDR within individual mass bins, and these binned data are subsequently fitted with broken power-law profiles (black solid lines; Equations~\ref{eq:fit_Mbh} and~\ref{eq:fit_Mgal}). TDR rises with increasing mass and exhibits a turnover above the characteristic break masses.}
    \label{fig:TDR-Mass}
\end{figure*}

{\noindent}We further investigate how the TDR correlates with the host galaxy mass and the associated black hole mass. Since the number of galaxies at any specific redshift is small and their mass ranges are limited, we combine all the galaxies across all the redshifts for this analysis. This is shown in Figure \ref{fig:TDR-Mass}, where the left and right panel represent the TDR as a function of $M_{\rm BH}$ and $M_{\rm gal}$, respectively. The TDEs from the primary host galaxies are marked as stars at all the redshifts, whereas the smaller galaxies (more on this is discussed in Section \ref{sec:ON}) are marked as points. 

To quantify the trends, following the method of \citet{Chang24}, we equally bin the $M_{\rm BH}$ ranges on a logarithmic scale (shown as the gray-shaded regions) and then calculate the linear average of TDR within each mass bin. The error bars represent the standard error on the mean TDR at each mass bin. We adopt a linear averaging scheme rather than, for example, a geometric mean, because this approach places more weight on systems with higher TDE rates. Since higher-rate systems are more likely to be detected observationally, this weighting is appropriate when comparing our theoretical predictions with observed event rates.  We note that this is intended as a descriptive tool rather than a rigorous population estimate. Nonetheless, we recover a characteristic trend, namely that the TDR increases with $M_{\rm BH}$ in the IMBH mass regime and then turns over and declines at higher $M_{\rm BH}$ in the SMBH regime.

We fit the averaged TDR as a function of $M_{\rm BH}$ using a broken power-law function as follows:
\begin{equation} \label{eq:fit_Mbh}
    \rm{TDR} =
    \begin{cases}
        10^{-4}\left(\frac{M_{\rm BH}}{M_b}\right)^{1.7} ~\rm{gal}^{-1}\rm{yr}^{-1}~~~; M_{\rm BH} < M_b, \\
        10^{-4} \left(\frac{M_{\rm BH}}{M_b}\right)^{-0.4}~\rm{gal}^{-1}\rm{yr}^{-1}~; M_{\rm BH} \geq M_b,
    \end{cases}
\end{equation}
where $M_b=1.8 \times 10^5M_\odot$ is the BH mass at which TDR turns around.

Intriguingly, similar trends -- an increasing TDR in the IMBH regime and a turnover of the TDR around $10^{5-6} M_{\odot}$ -- have also been found for TDEs in the local Universe in recent studies \citep{Polkas24, Chang24, Hannah24}, albeit with different normalization, exact turnover BH masses, and power-law slopes.  Likewise, when focusing solely on SMBHs, the overall decreasing trend of TDR with increasing $M_{\rm BH}$ broadly agrees with the previous work \citep{WangMerritt04, Stone16, Pfister20}. This general consistency may suggest that high-redshift and low-redshift galaxies share similar central structures, although further investigation is needed to substantiate this conclusion.

\begin{figure}
    \includegraphics[width=8.5cm]{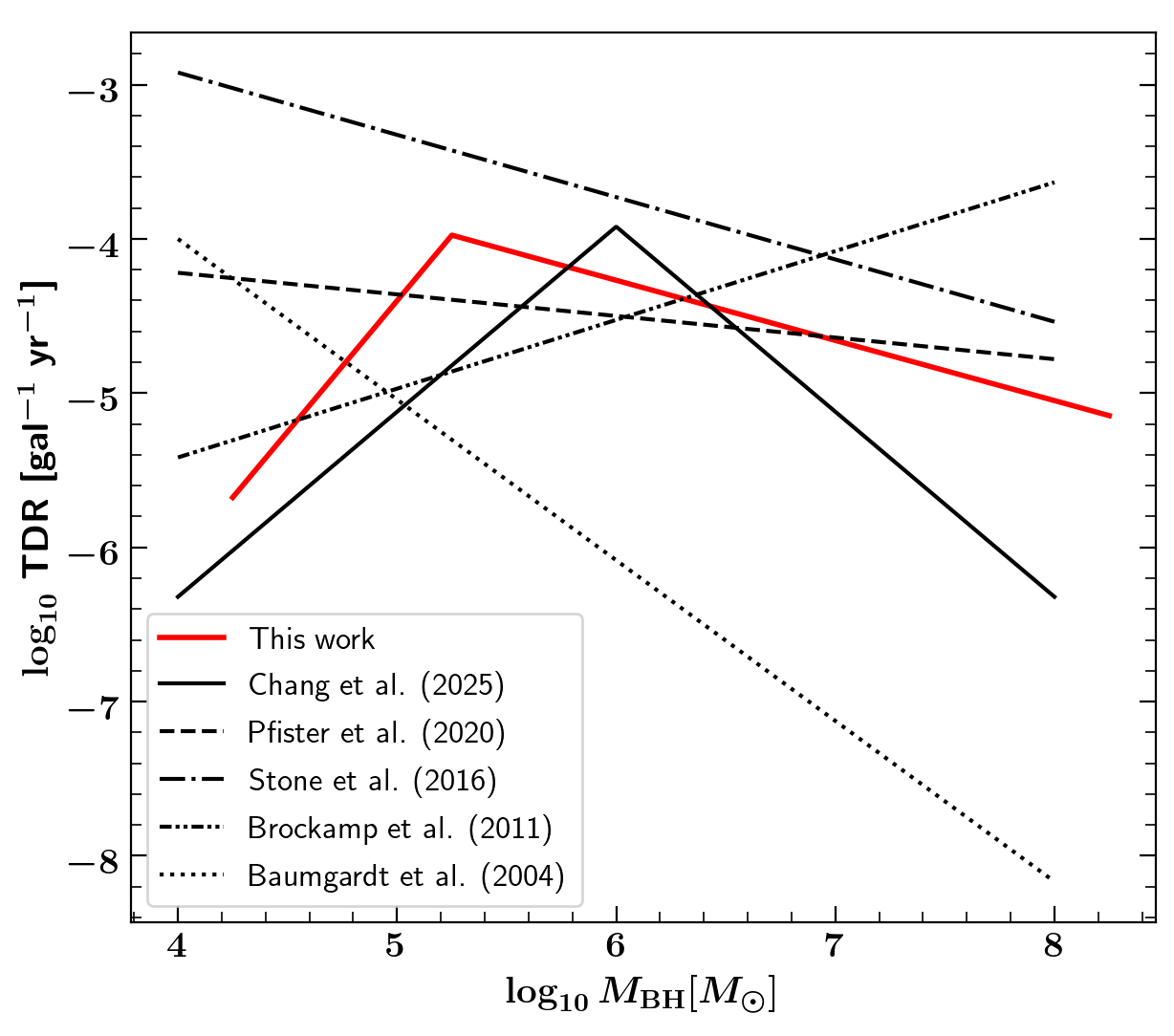}
    \caption{Comparison of the TDR-$M_{\rm BH}$ correlation obtained from this work with previous studies.}
    \label{fig:comp_all}
\end{figure}

For completeness, we present a summary of TDE rates as a function of $M_{\rm BH}$ estimated in this work alongside those from previous studies \citep{Baumgardt04, Brockamp11, Stone16, Pfister20, Chang24} in Figure \ref{fig:comp_all}. We caution, however, that these studies employ very different samples -- some simulated and some observed -- and are mostly restricted to low-redshift galaxies, whereas our work focuses on high-redshift systems.

The differences in the fitted parameters between this work and previous studies arises from several factors: e.g., using the observed versus simulated galaxy stellar profiles, different BH mass measurement techniques, the presence or absence of dense nuclear star clusters, and -- most importantly -- the distinct redshift ranges probed. Most of the previous studies have used the BH samples that are limited to the local Universe, while this work focuses on samples at $z \ge 1$. Hence, differences in the scaling relations are expected. However, qualitative agreement on the overall trend of the TDR with the $M_{\rm BH}$ and $M_{\rm gal}$ between this work $(z=1-10)$ and the previous studies (in the local Universe) hints at similar galaxy structures and BH-galaxy scaling across the cosmic time.

Likewise, the right panel of the Figure~\ref{fig:TDR-Mass} shows that the trend between the averaged TDR and $M_{\rm gal}$ is similar to that seen for the $M_{\rm BH}$. Specifically, the TDR rises with the $M_{\rm gal}$ for small galaxies, peaks at $M_g = 5 \times 10^9 M_{\odot}$, and gradually declines at the higher masses. This trend can be fit to a function as follows:
\begin{equation} \label{eq:fit_Mgal}
    \rm{TDR} =
    \begin{cases}
     10^{-4}\left(\frac{M_{\rm gal}}{M_g}\right)^{0.6} ~\rm{gal}^{-1}\rm{yr}^{-1}~~~; M_{\rm gal} < M_g, \\
     10^{-4} \left(\frac{M_{\rm gal}}{M_g}\right)^{-0.2}~\rm{gal}^{-1}\rm{yr}^{-1}~;  M_{\rm gal} \geq M_g,
    \end{cases}
\end{equation}

The overall trend is consistent with Figures \ref{fig:scaling1_z} and \ref{fig:scaling2_z}, where the TDR increases at high $z$ before reaching a peak at $z=2.5$ and gradually decreases thereafter. We investigate in more detail the impact of the properties of host galaxy stellar profiles on the associated TDR in the next section.

\subsection{Impact of Host Galaxy Stellar Profiles on TDR} \label{sec:TDR-host props}

\begin{figure*}
    \centering
    \begin{tabular}{c}
    \includegraphics[width=17cm]{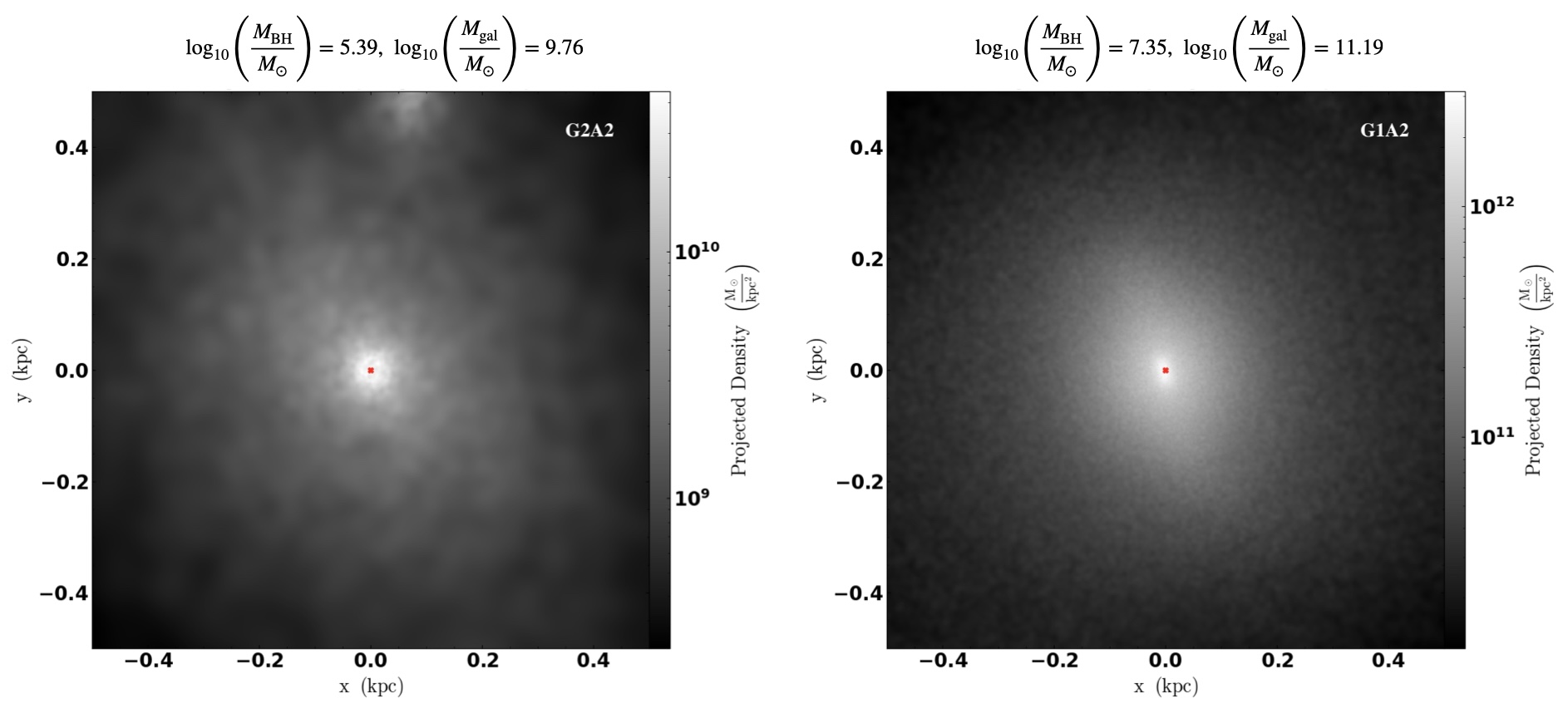}
    \end{tabular}
    \caption{The projected stellar densities around the BHs at the centers of two example galaxies, where the TDR is found to be highest in Figure \ref{fig:sfr}, marked as `G2A2' (left panel) and `G1A2' (right panel). The central BHs are shown as red crosses, and the corresponding $M_{\rm BH}$ and $M_{\rm gal}$ values are mentioned at the top of each panel. The color bars represent projected stellar densities.}
    \label{fig:sfr_check}
\end{figure*}

\noindent We begin with Figure~\ref{fig:sfr_check} that shows the projected stellar density distributions of G1A2 and G2A2 galaxies, associated with the highest TDR (Figure \ref{fig:sfr}). We find a dense stellar component in G1A2, which has a higher galaxy mass ($M_{\rm gal} \sim 10^{11} M_{\odot}$) and a central SMBH ($M_{\rm BH} \sim 10^7 M_{\odot}$). In contrast, G2A2 is a smaller galaxy ($M_{\rm gal} \sim 5 \times 10^9 M_{\odot}$) hosting an IMBH ($M_{\rm BH} \sim 10^5 M_{\odot}$) with the lower central stellar density. This motivates us to further investigate the stellar profiles of our galaxy sample and their impact on the TDR. 

\begin{figure*}
    \centering
    \includegraphics[width=16cm]{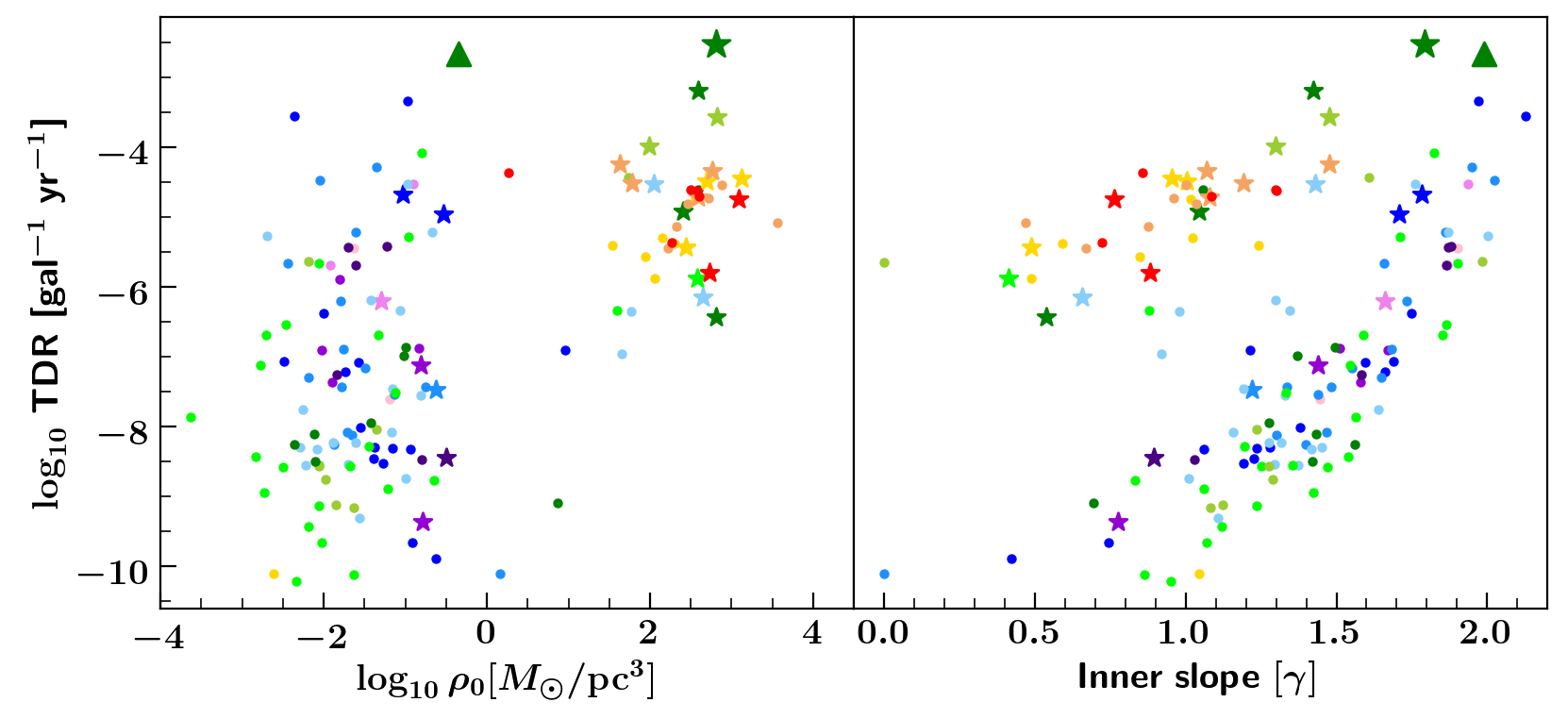}
    \caption{The correlated properties of the TDR vs the fitted parameters of stellar profiles at different redshifts. Left panel: correlation between TDR and stellar densities at the scale radius $(\rho_0)$. Right panel: correlation of TDR with the inner slope of the stellar profiles $(\gamma)$. The color scheme is the same as in Figure \ref{fig:TDR-Mass}, where different colors represent different redshifts. 
    Star symbols show the primary host galaxies at each $z$. Additionally, we mark G1A2 and G2A2 with a larger green star and a triangle, respectively. TDR appears to be correlated with both $\rho_0$ and $\gamma$, with a distinct evolution of these properties noticed with redshift. We further observe very steep $\gamma$ in both G1A2 and G2A2, which can contribute to the enhanced TDR in these systems. This is further discussed in Section \ref{sec:TDR-host props}.}
    \label{fig:fit_check}
\end{figure*}

Figure \ref{fig:fit_check} shows the correlation of the TDR with the inner stellar density (left panel) and the inner slope (right panel) of the simulated galaxy sample used in this work, with G1A2 and G2A2 highlighted. As expected, the TDR generally increases with the higher $\rho_0$ due to a richer stellar population available for disruption. Steeper inner slopes also enhance the TDR by driving the stars into the loss cone \citep{Chang24}. The highest TDR in G1A2 (star symbol) results from a combination of the high $\rho_0$ and steep $\gamma$, whereas in G2A2 (triangle symbol), a very steep $\gamma$ alone yields an elevated TDR despite its relatively lower $\rho_0$.

Furthermore, a closer look at the Figure~\ref{fig:fit_check} reveals the distinct redshift trends in both panels. The inner stellar densities of galaxies grow with decreasing redshift, with the most-dense systems found at $z \lesssim 4$. The right panel further shows a bimodality in the inner power-law slope $\gamma$: it increases at the early times until $z \sim 4$, after which the trend reverses and the inner stellar profiles become shallower at the lower redshifts.

\begin{figure*}
    \centering
    \includegraphics[width=18cm]{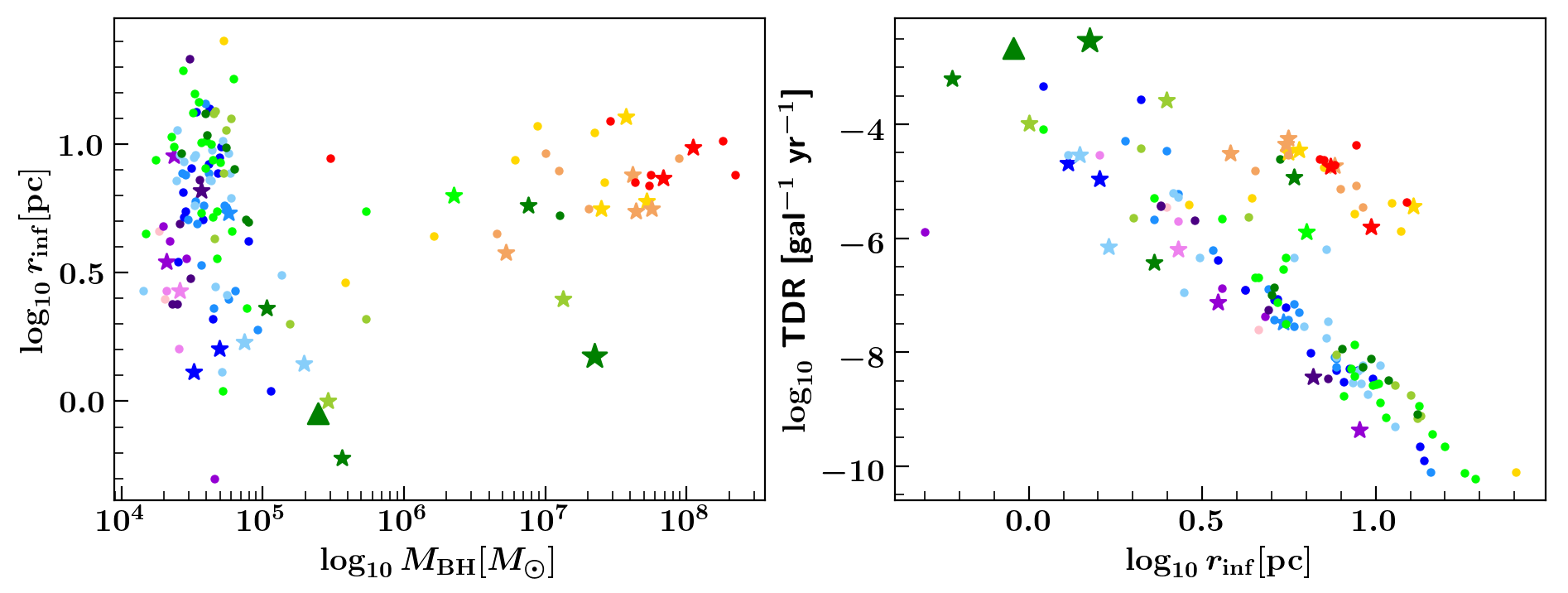}
    \caption{The correlation of the radius of influence $(r_{\rm inf})$ with the BH mass (left panel) and the TDR (right panel) at $z=1-10$. The symbols and the color schemes are the same as in Figure \ref{fig:TDR-Mass}. We note that $r_{\inf}$ decreases with redshift in the early Universe $(z \gtrsim 2)$. Moreover, $r_{\rm inf}$ strongly correlates with TDR at all redshifts, with the highest TDR produced at the smallest $r_{\rm inf}$. More is discussed in Section \ref{sec:TDR-host props}.}
    \label{fig:inf}
\end{figure*}

We further examine the black hole radius of influence ($r_{\rm inf}$), which significantly affects the TDR (see Section~\ref{sec:lc}). We note from the left panel of Figure \ref{fig:inf} that the IMBHs dominate at high redshift. This could be a result of the bursty star formation activity in the FIRE-2 that depletes the available gas reservoir needed for the BHs growth at high $z$ \citep{Angles17}. The increase of the stellar mass while keeping the $M_{\rm BH}$ almost constant (Figure 2 of \cite{Angles17}) explains the decreasing $r_{\rm inf}$ at $z \gtrsim 2$ in the left panel of Figure \ref{fig:inf}. Correspondingly, a rise in the TDR in this epoch due to the boosted stellar mass and the rapid SFR is noted from the right panel. Star formation becomes steady below $z \lesssim 2$, and the BHs grow rapidly to become more massive, leading to increased $r_{\inf}$, seen from Figure \ref{fig:inf}, left panel. However, much larger galaxies hosting the SMBH will lower the ratio of $r_{\inf}/$effective radius of the galaxy bulges, which disperse the stars, reducing the TDR in these systems \citep{Chang24}, which is seen from the right panel (yellow and red colors). This claim is also supported by the shallower inner slope of the galaxies at low redshift in Figure \ref{fig:fit_check}.

\begin{figure}
    \includegraphics[width=8.5cm]{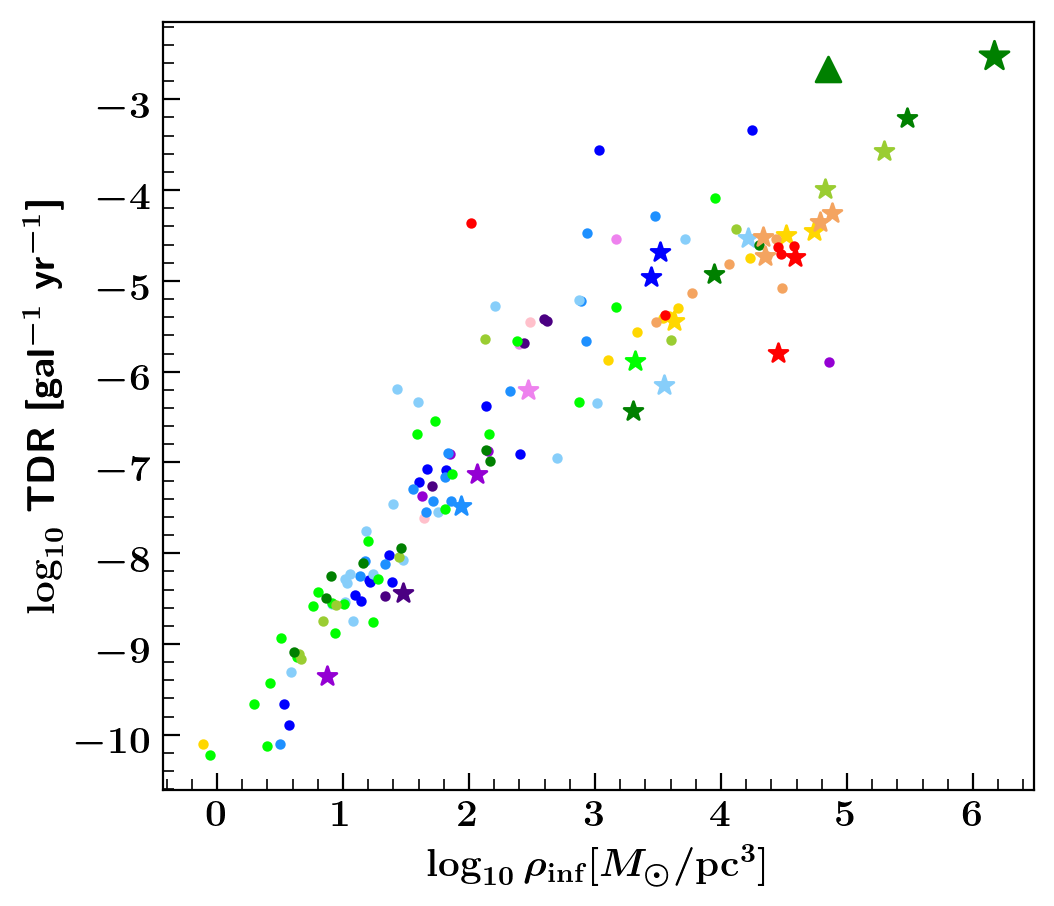}
    \caption{The strong correlation of the TDR with the stellar density at the influence radius $(\rho_{\rm inf})$ across $z=1-10$. The symbols and the color schemes are the same as in Figure \ref{fig:TDR-Mass}. We note that TDR strongly correlates with $\rho_{\rm inf}$ at all redshifts, with the highest TDR produced by the largest $\rho_{\rm inf}$.}
    \label{fig:rho_inf}
\end{figure}

Finally, in Figure \ref{fig:rho_inf}, we show the connection of the TDR with $\rho_{\rm inf}$, which is the stellar density at the influence radius. A strong correlation between the two parameters is noticed, consistent with the previous studies \citep{Pfister20, Chang24}. This can be naturally explained as a consequence of the highest stellar flux at the critical radius, which is the same as $r_{\inf}$ for $M_{\rm BH}< 10^8M_\odot$, as discussed in Section \ref{sec:lc}. We also find a large influence density in both the G1A2 and G2A2, explaining the surge in the TDR in these galaxies.

\subsection{TDEs in Satellite Galaxies} \label{sec:ON}

\begin{figure}
    \includegraphics[width=8.5cm]{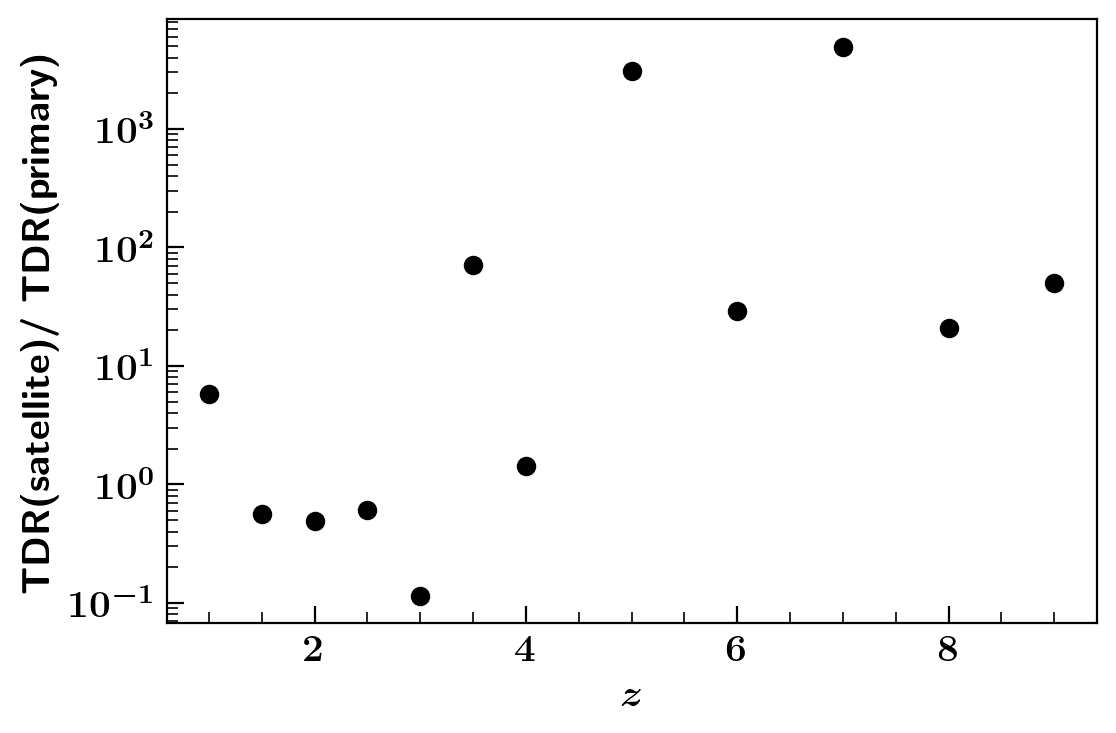}
    \caption{The ratio of the total TDR from all satellite galaxies to that from primary galaxies at each redshift. The TDE contribution from satellite galaxies is significantly higher in the early Universe.}
    \label{fig:onfrac}
\end{figure}

\begin{figure}
    \includegraphics[width=8.5cm]{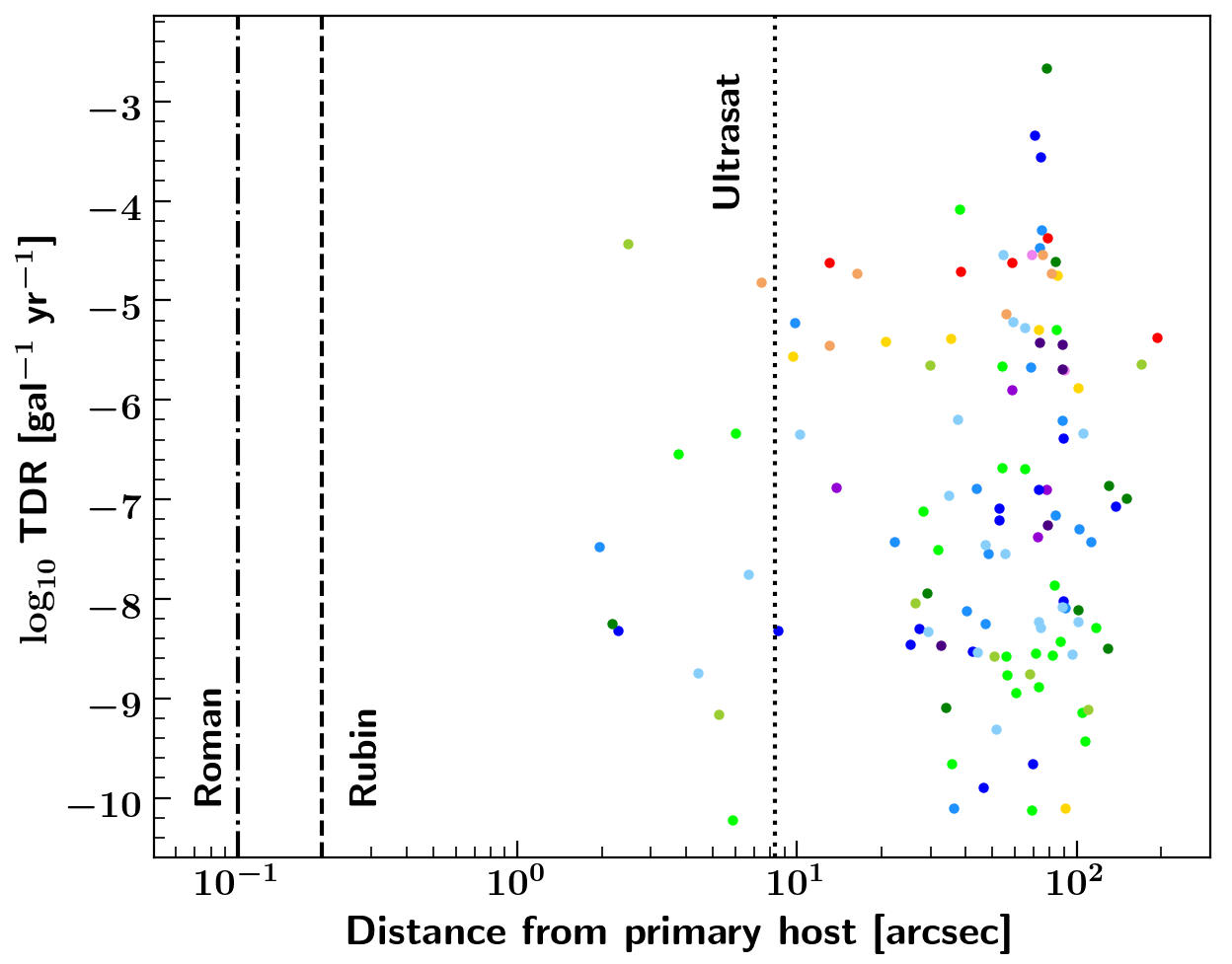}
    \caption{The TDR of satellite galaxies versus their offset distances from the primary hosts at each redshift. Black lines mark the angular resolution limits of Roman (dashed-dotted), Rubin (dashed), and ULTRASAT (dotted), respectively. While there is a spread in TDR among satellite galaxies, the majority of TDEs are located at angular separations well within the resolution limits of all three telescopes. Colors are consistent with the redshift scheme used in Figure~\ref{fig:TDR-Mass}.}
    \label{fig:on}
\end{figure}
\noindent
In this section, we focus on examining the TDEs in the satellite galaxies, defined as all the nonprimary galaxies in our simulated galaxy sample. Because these satellite galaxies are offset from the primary (most-massive) galaxy in the cluster, TDEs occurring within them may be observed as the off-nuclear TDEs, depending on the satellite's size and the distance. Detecting such TDEs is crucial for identifying IMBHs in the globular clusters and dwarf galaxies, and for tracing the galaxy merger histories across the cosmic time.

As shown in Figure~\ref{fig:TDR-Mass}, satellite galaxy TDRs can be as high as those of the primary galaxies, spanning a broad range from $10^{-10}$ to $10^{-3} \, \text{gal}^{-1} \, \text{yr}^{-1}$ across all redshifts. In Figure~\ref{fig:onfrac}, we present the fraction of the total TDR from the satellites versus the primary galaxies as a function of $z$. It can be seen that the satellite TDR fraction increases substantially at high redshifts, highlighting their importance for studying the IMBHs and galaxy merger histories in the early cosmos.

Figure~\ref{fig:on} shows the TDR of the individual satellite galaxies as a function of the distance from their primaries, with the transverse physical distances converted to the angular sizes in the local Universe.  
The angular resolutions of Roman, Rubin, and ULTRASAT \citep{Shvartzvald24} are marked in the figure. Notably, almost all the satellite TDEs in our sample can be resolved by these missions in most cases. Nevertheless, the actual detectability depends on the additional factors, including the TDE luminosity, instrument flux limits, and survey specifications (e.g., field of view and limiting magnitude). A detailed analysis is beyond the scope of this work, and we defer a thorough investigation to a future study.

\section{Summary and Discussions} \label{sec:summary}
\noindent In this work, we present a comprehensive analysis of how TDE rates evolve over cosmic history spanning a wide redshift range of $1 < z < 10$, based on a sample of simulated galaxies drawn from the high-resolution cosmological zoom-in simulation FIRE-2.
The simulation covers a diverse population of black holes and galaxies, ranging from the SMBHs in primary hosts to IMBHs in satellite galaxies, which allows us to study how TDE rates correlate with host galaxy properties and their redshift evolution. 
Below we summarize the key findings of this paper: 
\begin{enumerate}
    \item Whereas current TDR calculations are largely confined to the local Universe, we here predict, for the first time, the redshift evolution of the TDR across $z=1-10$ (Figure \ref{fig:ratez}) and provide an estimation of this evolution in the early Universe (Equation \ref{eq:ratez}).

    \item We observe a strong correlation of the TDR with the global SFR of their host galaxies (Figure \ref{fig:sfr}) and estimate a scaling relation between TDR and SFR across the entire redshift range (Equation \ref{eq:TDR-SFR}). We find that TDR closely follows the SFR, where both increase with time in the early Universe before reaching a peak at $z\sim 2.5$, and moderately decline afterwards.
    
    \item We find that TDR correlates well with both $M_{\rm BH}$ and $M_{\rm gal}$ for the overall galaxy sample (Figure~\ref{fig:TDR-Mass}, Equation~\ref{eq:fit_Mbh}, Equation~\ref{eq:fit_Mgal} ). Specifically, TDR increases with BH mass in the IMBH regime, peaks at $M_{\rm BH} \sim 10^5 M_{\odot}$ (and $M_{\rm gal} \sim 10^9 M_{\odot}$), and then declines at the high-mass end. This trend, previously seen in local galaxies, intriguingly persists at high redshifts. 

    \item We also examine various components of the stellar distributions in galactic centers at different redshifts in connection to their associated TDE rates, shown in Figure \ref{fig:sfr_check}, \ref{fig:fit_check}, \ref{fig:inf} and \ref{fig:rho_inf}. The combined findings from these analyses hint that TDR is large in galaxies with high inner density $(\rho_0)$, steep inner slope $(\gamma)$, small influence radius $(r_{\inf})$ and large stellar density at this scale $(\rho_{\inf})$. 

    \item Finally, we examine the detectability of TDEs in satellite galaxies. The fraction of TDEs originating from satellite galaxies increases significantly at high redshifts, underscoring their potential as probes of IMBHs and galaxy assembly in the early Universe. Encouragingly, most of these events fall within the angular resolution limits of the facilities such as Roman, Rubin, and ULTRASAT, indicating that satellite TDEs are detectable with appropriate survey strategies.

\end{enumerate}

Recent studies have explored the redshift evolution of the TDR, based on theoretical models \citep{Polkas24, Melchor25, Karmen26}. We note that drawing firm quantitative conclusions on TDR evolution with redshift using current cosmological simulations faces several limitations. First, feedback physics such as AGN feedback, which is not included in the Massive Halo suite, can affect host galaxy properties \citep{Dubois16, rkc20, rkc22, Byrne24, KV24}. Its absence may artificially enhance SFRs and central stellar densities, thereby boosting TDE rates, particularly at $z \lesssim 2$ \citep{Wellons20, Parsotan21, Cochrane23}. On a more positive note, our sample is dominated by lower-mass BHs ($M_{\rm BH} \lesssim 10^6 M_{\odot}$), for which AGN feedback is negligible.  Second, the finite resolution of FIRE-2 -- although the best currently available -- still prevents us from directly resolving nuclear star clusters, which are known to substantially boost local TDRs. Third, because the FIRE-2 simulation was evolved only to $z=1$ to mitigate overcooling \citep{Wetzel23}, our TDR constraints do not extend to $z=0$. Moreover, we note that, being zoom-in simulations, the halo mass range is narrow in the Massive Halo suite. This limits the total number of galaxies available for TDE analysis. However, the combined numbers of all the galaxies across $z=1-10$ is significantly larger compared to previous studies. Finally, we note that our analysis assumes spherical symmetry for the stellar distribution -- a common simplification -- which may affect the TDE rates. Further analytical or numerical work is needed to properly quantify these effects.

Despite these limitations, we demonstrate that constraining the cosmic evolution of TDR using cosmological simulations is a promising avenue, and we encourage future work to revisit these TDR calculations with next-generation simulations. Current and next-generation telescopes, including Roman, Rubin, and ULTRASAT, are expected to detect numerous TDEs across a broad redshift range, providing an ideal test bed for comparing our predictions with observations. Such comparisons will allow us to validate and refine models of TDE rates, constrain the interplay between black hole growth, stellar dynamics, and galaxy evolution across cosmic time, and ultimately establish TDEs as a reliable probe of the coevolution of black holes and their host galaxies from the early Universe to the present day.

\begin{acknowledgments}
We thank the referee for many important suggestions to better shape the manuscript. The authors thank Prof. Sarah Wellons for providing valuable comments that helped improve the manuscript. We thank C. Bottrell, S. Ji, P. Natarajan, and F. Yuan for useful discussions. R.K.C., L.D. and J.C. acknowledge support from the National Natural Science Foundation of China and the Hong Kong Research Grants Council (NSFC/RGC JRS N\_HKU782/23, RGC GRF 17314822). R.K.C. would like to acknowledge the financial support provided by the Anusandhan National Research Foundation (ANRF), a statutory body of the Department of Science and Technology (DST), Government of India, through the National Post-Doctoral Fellowship (NPDF) (grant No. PDF/2025/004682).
\end{acknowledgments}

\appendix
\counterwithin{figure}{section}

\section{Stellar Densities and Radial Profiles in FIRE-2} \label{app:rate_eq}

\begin{figure*}
    \centering
    \begin{tabular}{c}
    \includegraphics[width=9cm]{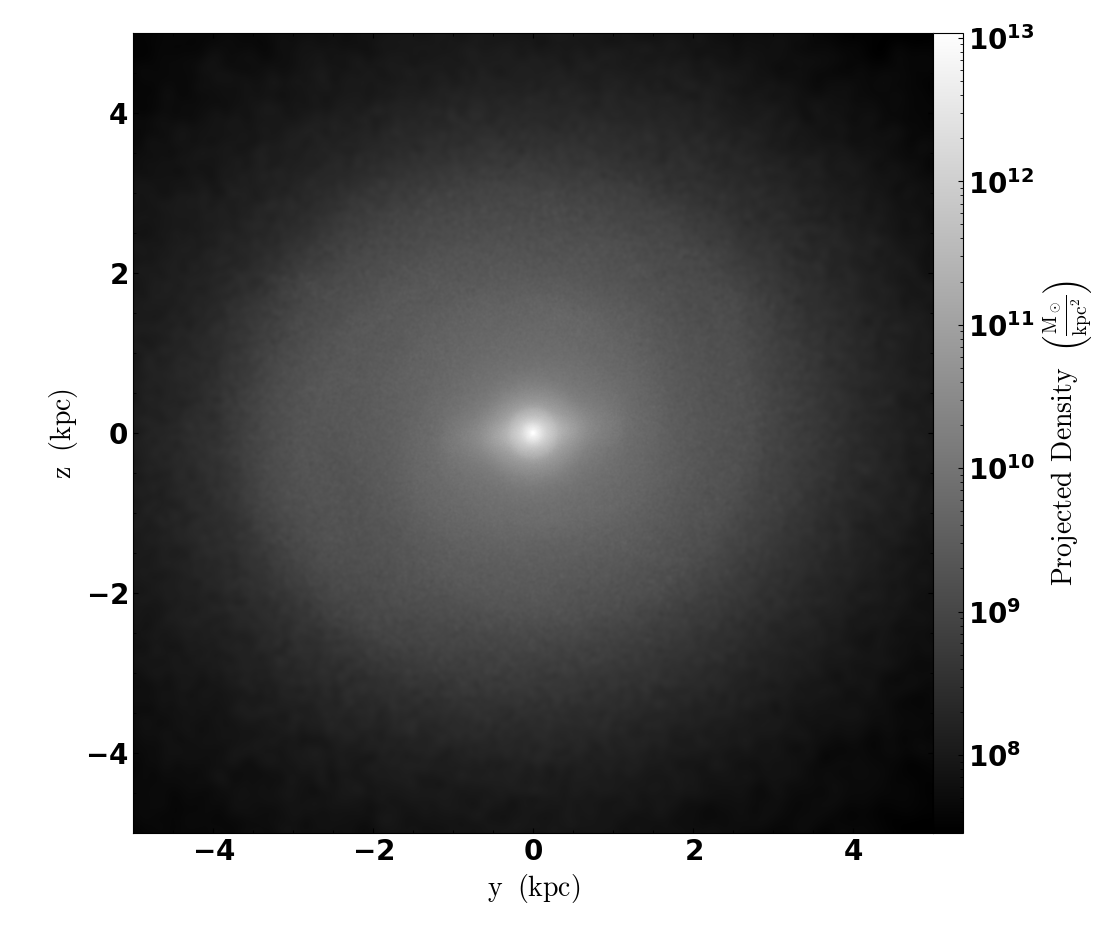}
    \includegraphics[width=9cm]{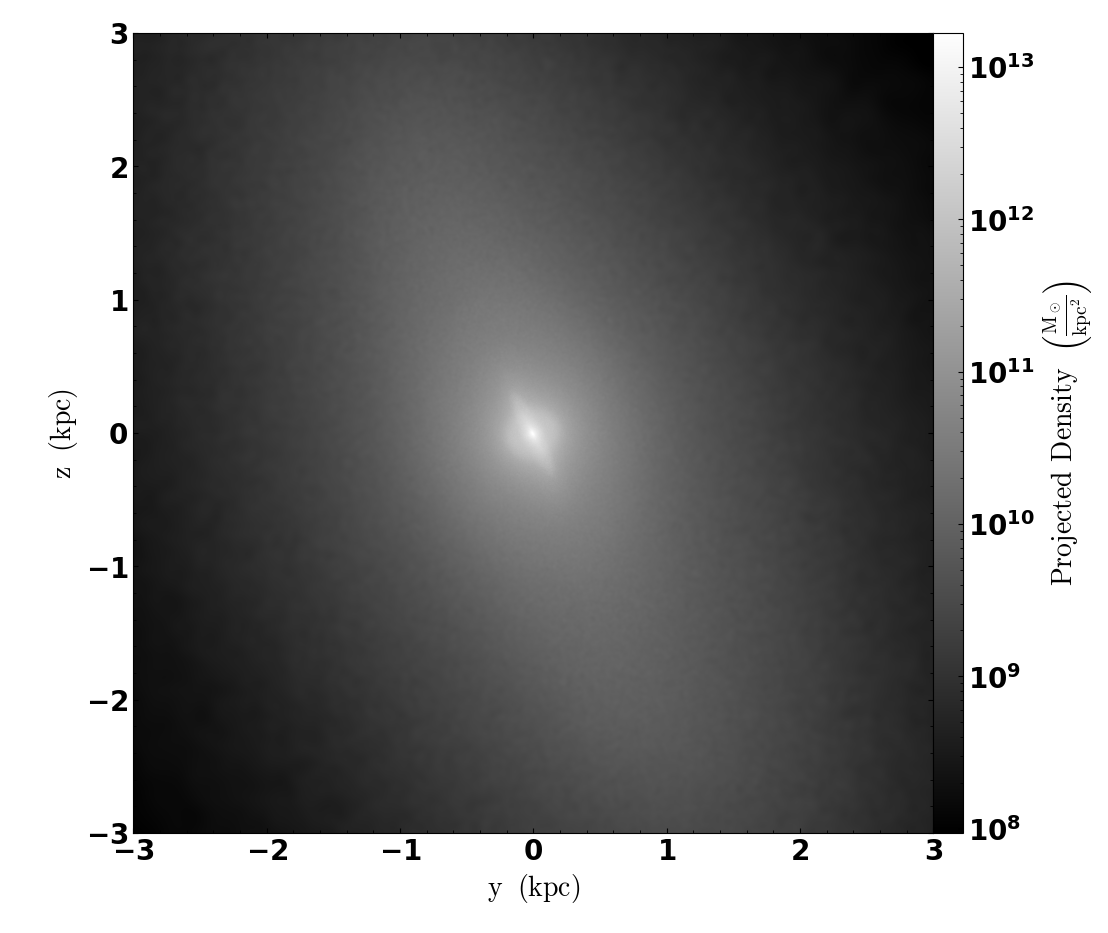}\\
    \includegraphics[width=9cm]{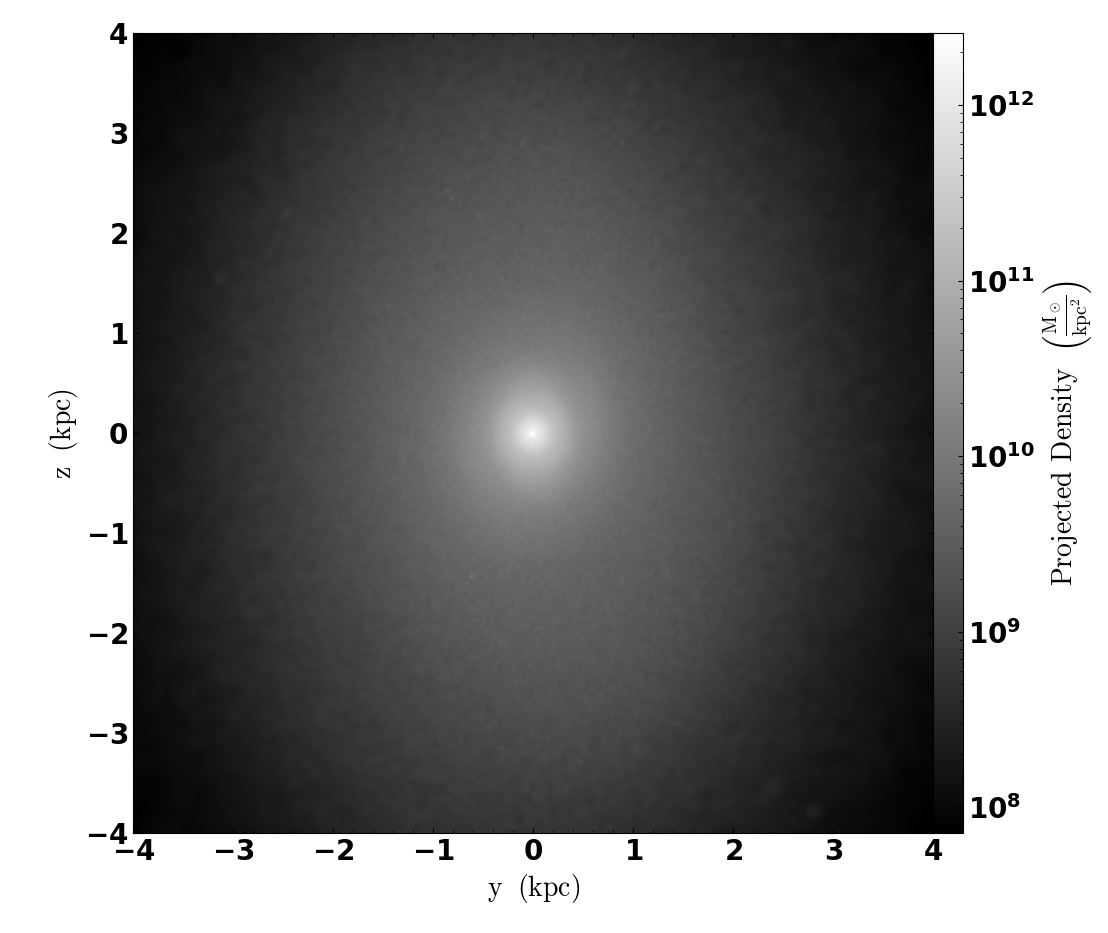}
    \includegraphics[width=9cm]{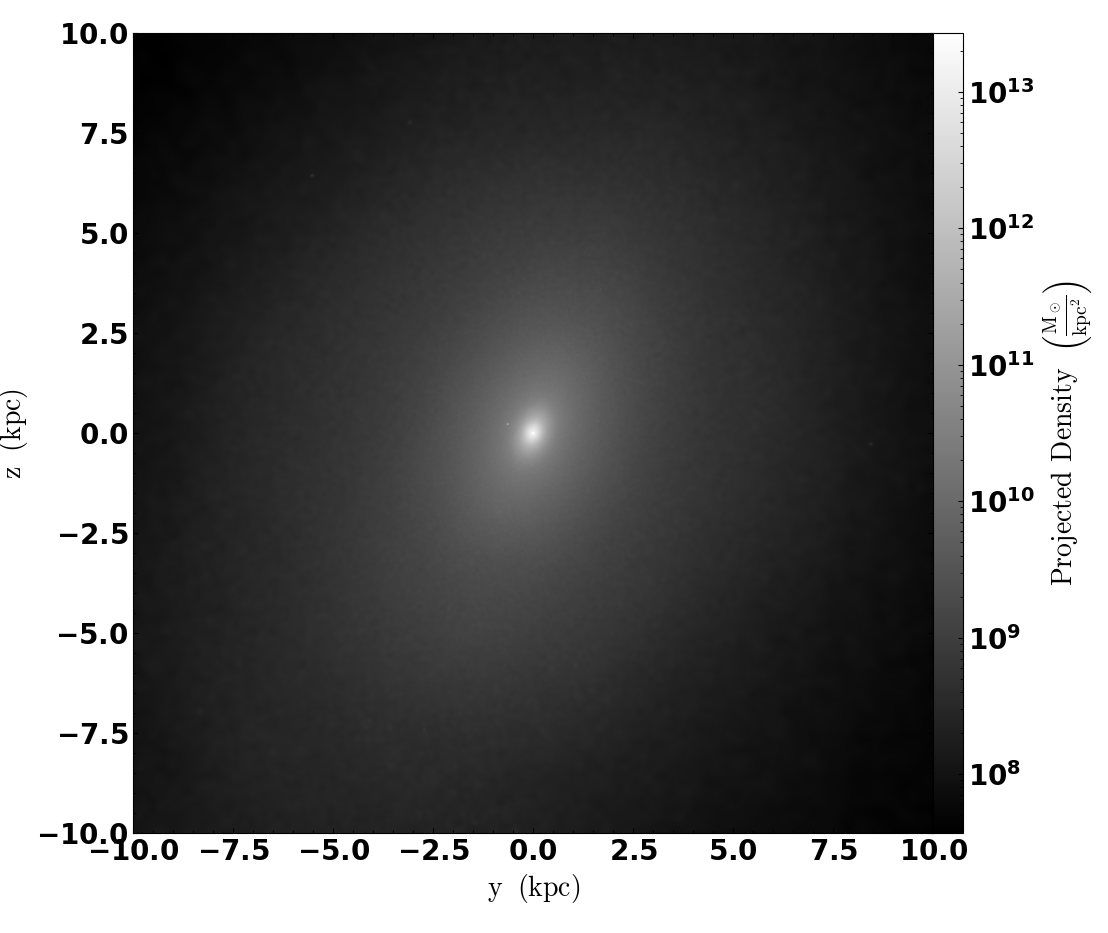}
    \end{tabular}
    \caption{Stellar densities within the primary host galaxies at $z=1$ in A1 (top left), A2 (top right), A4 (bottom left), and A8 (bottom right) runs, projected along the x-axis. The choice of such projection is simply because of viewing purposes as all four galaxies are viewed face-on in this orientation. Color bars represent the projected densities in each panel.}
    \label{fig:maps}
\end{figure*}

\begin{figure}
    \centering
    \includegraphics[width=17cm]{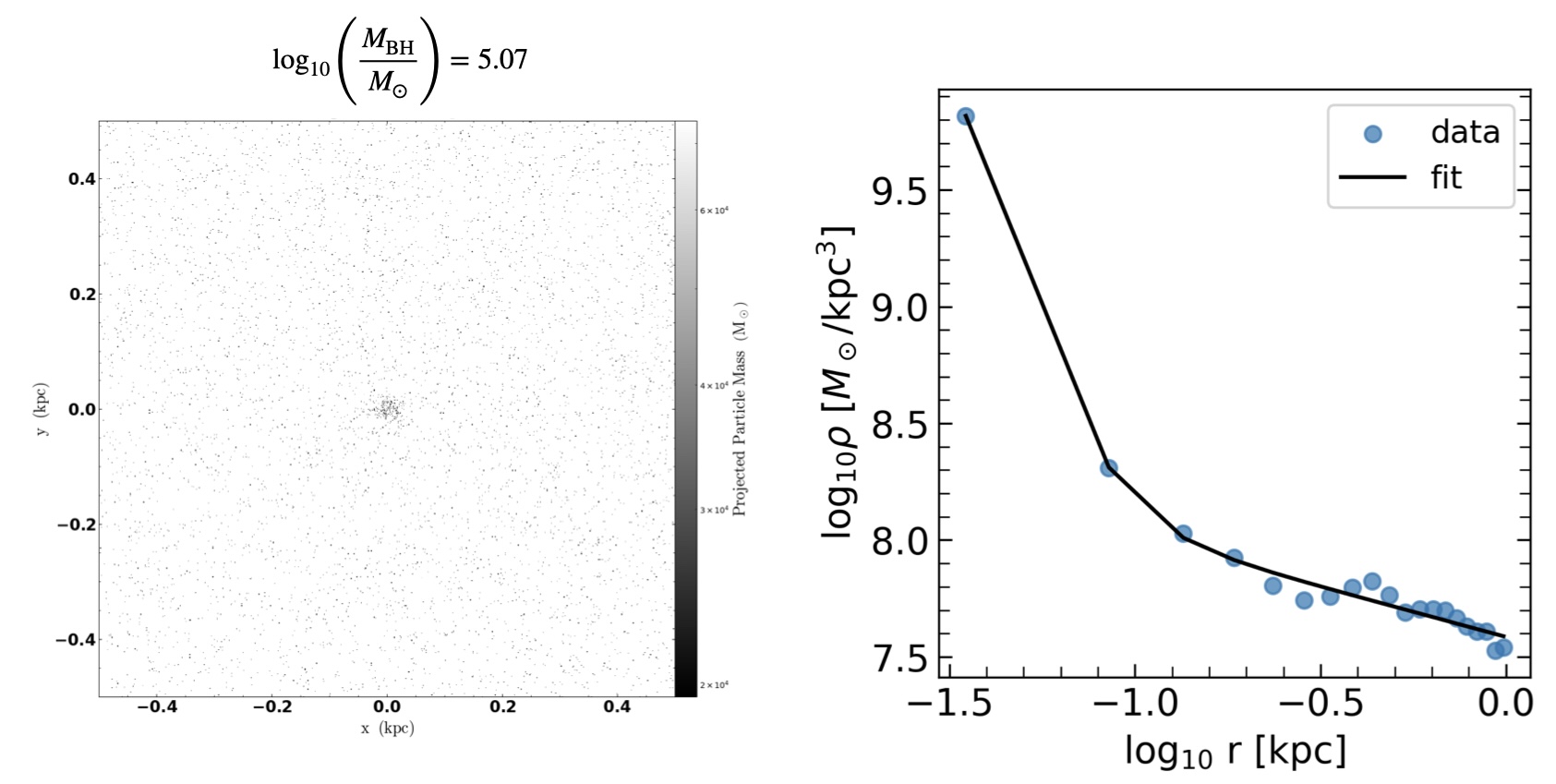}
    \caption{The distribution of stars (left panel) and the corresponding radial density profile (right panel) at the center of a small galaxy at $z=1$ in the A1 run. In the left panel, the color bar shows the particle mass summed along the z-direction. Double power-law fit (Equation \ref{eq:dp_profile}) is shown as a black solid line in the right panel. It is to be noted that the inner power-law slope is steeper than the outer profile. Galaxies with this type of stellar profile are discarded from our sample. More details are discussed in Section \ref{subsec:sim_rate}.}
    \label{fig:app_stellar_density}
\end{figure}

\begin{figure*}
    \centering
    \begin{tabular}{c}
    \includegraphics[width=16cm]{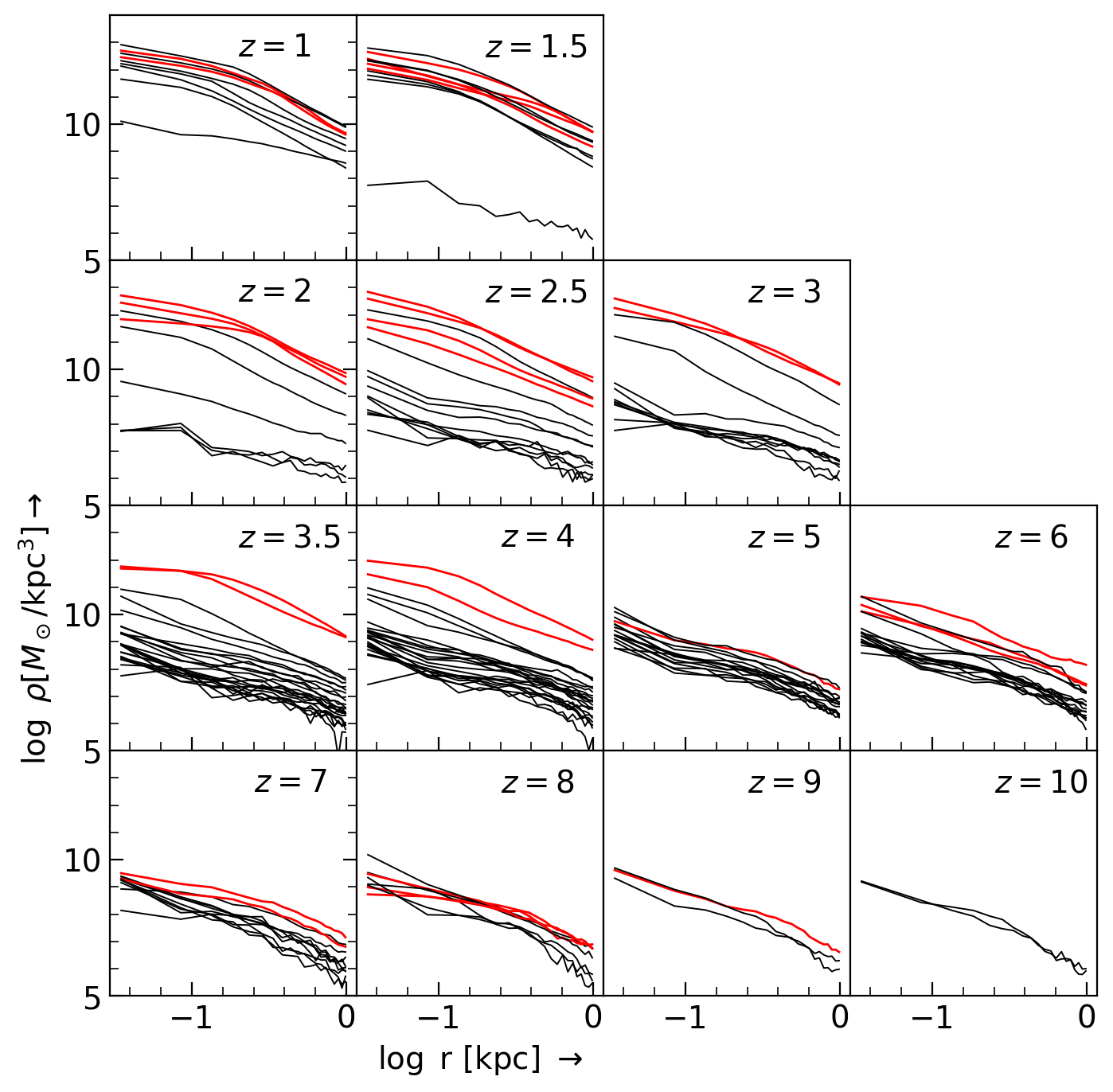}
    \end{tabular}
    \caption{The radial stellar density profiles of all the galaxies with $M_{\rm gal}>10^7M_\odot$ across $z=1-10$ in all four Massive Halo runs. In each panel, red solid lines represent the primary host galaxies at individual redshifts, whereas black solid lines represent the rest of the galaxies. We fit these with the double power-law profile (Equation \ref{eq:dp_profile}) and define our selected sample based on the fitted parameters (Section \ref{subsec:sim_rate}).}
    \label{fig:prof_all}
\end{figure*}

\begin{figure*}
    \centering
    \begin{tabular}{c}
    \includegraphics[width=16cm]{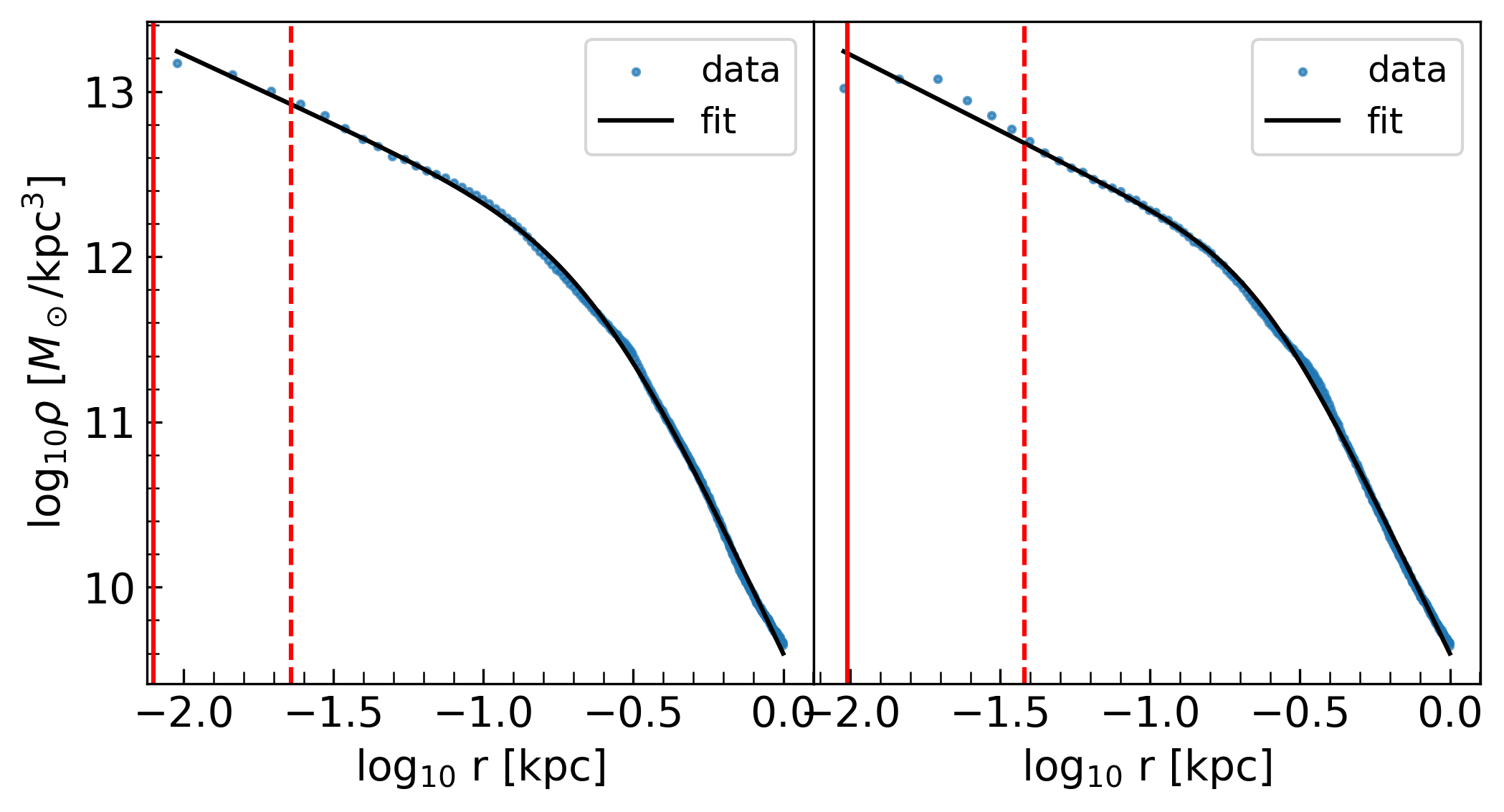}
    \end{tabular}
    \caption{Radial density profile of stars in the primary host galaxies of A1 (left) and A4 (right) run at $z=1$. The radial scale is extended up to the $r_{\rm inf}$ in both the cases (red solid lines) to examine the stellar distribution at that scale. Moreover, we mark the critical radius as the red dashed lines in both the panels.}
    \label{fig:pro_inf}
\end{figure*}

Figure \ref{fig:maps} shows the stellar densities in the most-massive galaxies at $z=1$ in the A1 (top left), A2 (top right), A4 (bottom left), and A8 (bottom right) Massive Halo suite. Furthermore, Figure \ref{fig:app_stellar_density} displays an example of the stellar distribution and the corresponding density profile with fitted parameters $\gamma > \delta$, the type of systems that are discarded from our sample. It can be noticed from the left panel that a few stellar particles are present at the center, causing a steep slope in the density profile, which could be a numerical artifact of the simulation. In Figure \ref{fig:prof_all} we show the radial stellar density profiles of all the galaxies with $M_{\rm gal} > 10^7 M_\odot$. We fit these profiles with double power-law functions (Equation \ref{eq:dp_profile}) and choose only those that satisfy our selection criteria, mentioned in Section \ref{subsec:sim_rate} $(\gamma< \delta,~ 0 < \gamma < 3)$. Finally, Figure \ref{fig:pro_inf} shows stellar density profiles in the primary galaxies in A1 and A4 runs at $z=1$, extended down to the scale of $r_{\rm inf}$ in these systems as shown from Figure \ref{fig:inf}. We have also marked the $r_{\rm crit}$ as red dashed lines in both the panels. It is noted that the difference between $r_{\rm inf}$ and $r_{\rm crit}$ is only a factor of $\sim 2$, supporting the validity of the assumption as mentioned in Section \ref{sec:lc}.

\section{Scaling Relations in Galaxies with Asymmetric Stellar Distribution} \label{app:asymmetric}

\begin{figure*}
    \centering
    \begin{tabular}{c}
    \includegraphics[width=16cm]{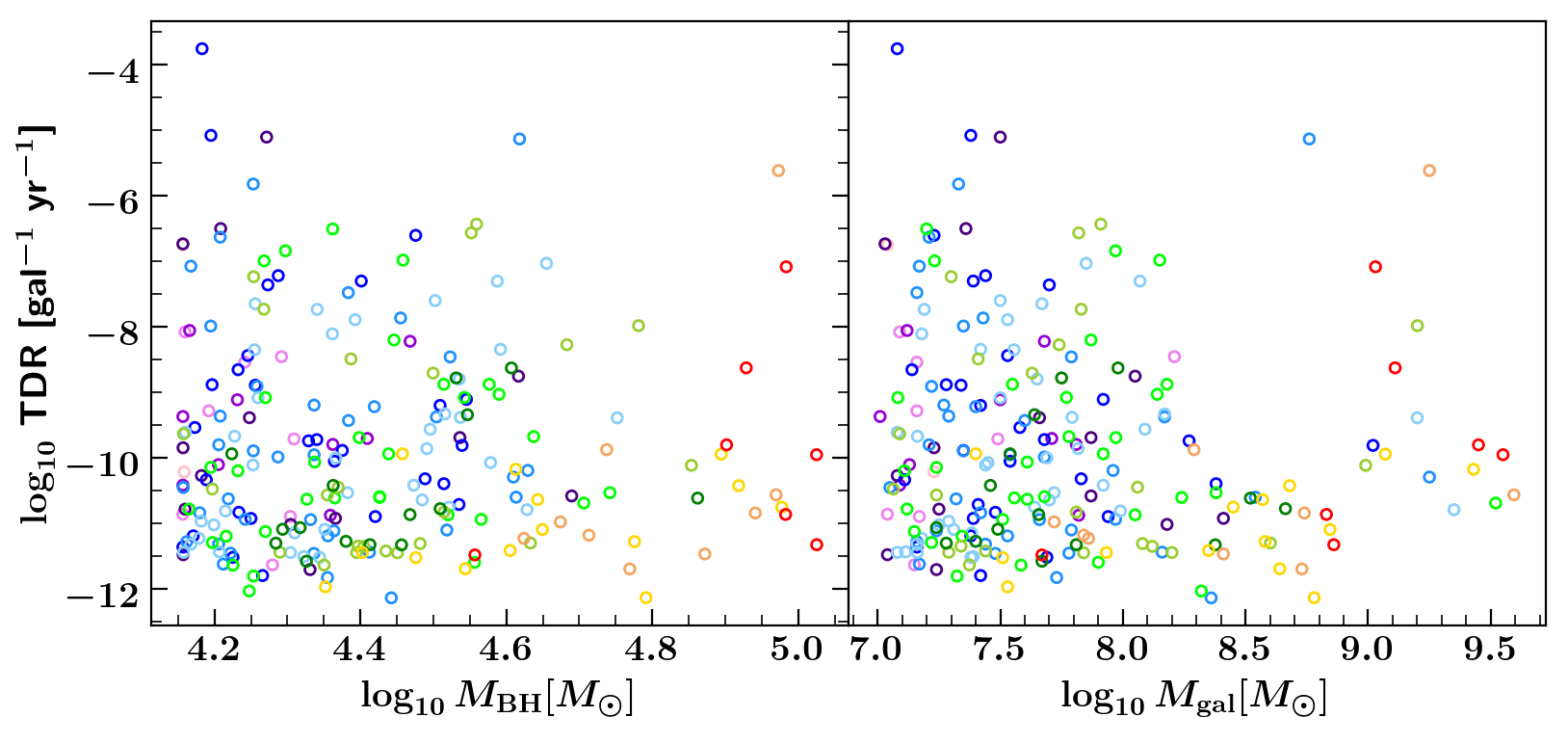}
    \end{tabular}
    \caption{Correlation of TDR with the mass of BH and host galaxies with visible asymmetric stellar distributions. The color scheme is the same as Figure \ref{fig:TDR-Mass}. No correlation is noticed from both of the panels.}
    \label{fig:scaling_asym}
\end{figure*}

We investigate in this section the correlation between the TDR, $M_{\rm BH}$ and $M_{\rm gal}$ for the galaxies in which the stellar distributions look asymmetric and are discarded from the final sample used in this paper. As discussed in Section \ref{sec:summary}, calculation of TDR using \textsc{phaseflow} assumes a spherical stellar distribution and a deviation from such a structure might result in inconsistencies in the estimated TDR. This is reflected in Figure \ref{fig:scaling_asym}. No correlation is noticed from both TDR-$M_{\rm bH}$ and TDR-$M_{\rm gal}$ in the asymmetric systems shown in this figure as opposed to the spherical systems as seen from Figure \ref{fig:TDR-Mass}. We also notice that the overall TDR is lower in the case of the asymmetric stellar distribution compared to the spherical systems. Hence, these discarded galaxies in our main analysis should not greatly affect the observed TDR.

\bibliography{ref}{}
\bibliographystyle{aasjournalv7}

\end{document}